\documentclass[10pt,conference]{IEEEtran}
\usepackage{amssymb, amsthm, amsfonts, latexsym, amsmath, times,balance,multicol}
\usepackage[pdftex]{graphicx}
\usepackage{algorithm, algorithmicx, graphics,mathtools,lipsum}
\usepackage{url, xspace, alltt, times, multirow,cite, floatflt,mathptmx}
\usepackage[dvipsnames]{xcolor}
\usepackage{caption}
\usepackage{subcaption}
\usepackage{verbatim}
\usepackage{grffile} 
\usepackage{morefloats}
\usepackage{float}
\usepackage{multirow, graphicx}
\usepackage{latexsym, dsfont, mathptmx, color, colortbl, mathptmx, bm}
\usepackage{stmaryrd}
\usepackage{caption}
\usepackage{xspace} 
\usepackage{verbatim}
\usepackage{grffile}
\usepackage{hyperref}
\DeclareMathAlphabet{\mathcal}{OMS}{cmsy}{m}{n}
\DeclareSymbolFont{largesymbols}{OMX}{cmex}{m}{n}
\DeclareMathOperator*{\argmax}{\mathbf{argmax}}

\newtheorem*{remark}{Remark}

\usepackage{soul}
\usepackage{cancel}
\usepackage[normalem]{ulem}
\usepackage{listings}
\usepackage{tabularx}
\usepackage{placeins}
\usepackage{color}
\usepackage[noend]{algpseudocode}
\usepackage{mathtools}
\usepackage[T1]{fontenc}
\definecolor{gray}{rgb}{0.4,0.4,0.4}
\definecolor{darkblue}{rgb}{0.0,0.0,0.6}
\definecolor{cyan}{rgb}{0.0,0.6,0.6}
\definecolor{backcolour}{rgb}{0.95,0.95,0.92}

\hypersetup{
    colorlinks=true,
    urlcolor=NavyBlue,
    linkcolor=black,
    citecolor=black
}

\lstdefinelanguage{XML}
{
  morestring=[b]",
  morestring=[s]{>}{<},
  morecomment=[s]{<?}{?>},
  stringstyle=\color{black},
  identifierstyle=\color{darkblue},
  keywordstyle=\color{cyan},
  morekeywords={xmlns,version,type}
}

\newtheorem{counter-example}{Counter Example}

\newcommand{\mathsc}[1]{{\normalfont\textsc{#1}}}

\newcommand{\Retired}{\ensuremath{\mathsf{Ins}}\xspace}
\newcommand{\phaseStartIns}{\ensuremath{\mathsf{StartIns}}\xspace}
\newcommand{\phaseEndIns}{\ensuremath{\mathsf{EndIns}}\xspace}
\newcommand{\Sched}{\ensuremath{\mathsf{Schedule}}\xspace}
\newcommand{\maxSlackTask}{\ensuremath{\mathsc{MaxSlackTask}}\xspace}
\newcommand{\finishTime}{\ensuremath{\mathsc{GetFinishTime}}\xspace}

\newcommand{\getSched}{\ensuremath{\mathsc{GetSchedTasks}}\xspace}
\newcommand{\resourceAlloc}{\ensuremath{\mathsc{ResourceAlloc}}\xspace}
\newcommand{\resetSegment}{\ensuremath{\mathsc{ResetSegment}}\xspace}
\newcommand{\computeRetiredInstruction}{\ensuremath{\mathsc{ComputeIns}}\xspace}
\newcommand{\shiftSuccessors}{\ensuremath{\mathsc{ReleaseSuccessors}}\xspace}
\newcommand{\done}{\ensuremath{\mathsf{Done}}\xspace}

\newcommand{\tnext}{\ensuremath{t_{\mathsf{next}}}\xspace}
\newcommand{\best}{\ensuremath{\mathsf{best}}\xspace}

\newenvironment{packeditemize}{
\begin{itemize}
  \setlength{\itemsep}{0.3pt}
  \setlength{\parskip}{2pt}
  \setlength{\parsep}{0pt}
}{\end{itemize}}

\def\sys{\mbox{\textsc{Cord}}\xspace}
\def\algo{\mbox{\textsc{Cord}}\xspace}
\def\baseline{\mbox{\textsc{Decomp}}\xspace}

\newcommand{\abby}[1]{{\color{red} \underline{\bf Abby says:} #1\xspace}}

\title{{\LARGE{\bf{\sys: Co-design of Resource Allocation and Deadline Decomposition with Generative Profiling}}}
}

\author{Robert Gifford, Abby Eisenklam, Georgiy Bondar, Yifan Cai, } 

\author{
    \IEEEauthorblockN{Robert Gifford\IEEEauthorrefmark{1}, Abby Eisenklam\IEEEauthorrefmark{1}, Georgiy A. Bondar\IEEEauthorrefmark{2}, Yifan Cai\IEEEauthorrefmark{1}, Tushar Sial\IEEEauthorrefmark{3}, Linh Thi Xuan Phan\IEEEauthorrefmark{1},\\Abhishek Halder\IEEEauthorrefmark{3}}
    \IEEEauthorblockA{\IEEEauthorrefmark{1}Department of Computer and Information Science, University of Pennsylvania,
    \\\{rgif, aei, caiyifan, linhphan\}@seas.upenn.edu}
    \IEEEauthorblockA{\IEEEauthorrefmark{2}Department of Applied Mathematics, University of California Santa Cruz
    \\gbondar@ucsc.edu}
    \IEEEauthorblockA{\IEEEauthorrefmark{3}Department of Aerospace Engineering, Iowa State University
    \\\{tsial, ahalder\}@iastate.edu}
} 

\begin{document}

\maketitle

\pagestyle{plain}
\thispagestyle{plain}

\begin{abstract}

As multicore hardware is becoming increasingly common in real-time systems, 
traditional scheduling techniques that assume a single worst-case execution 
time for a task are no longer adequate, since they ignore the impact of 
shared resources on execution time. 
When tasks execute concurrently on different cores, their execution times 
often vary substantially with their allocated budgets of shared resources,
such as cache and memory bandwidth. Even under a specific resource allocation, the resource use 
pattern of a task also changes with time during a job execution. It is therefore 
important to consider the relationship between multicore resources and execution time 
in task modeling and scheduling algorithm design.

In this paper, we propose a much more precise execution model for DAG-based real-time tasks that captures 
the time-varying resource use characteristics of a task under different budgets
of shared resources. We present a generative resource profiling algorithm that 
efficiently predicts, from limited measurement data, the resource profile  
of a task at any time during its execution under a given resource budget. 
The generative profiles can then be used to construct the execution
models for tasks, using which one can make informed resource allocation decisions.
We further introduce a multicore resource allocation and deadline decomposition 
co-design technique for DAG-based tasks that leverages the generated execution
models to jointly allocate resources and deadlines to subtasks, to 
maximize resource efficiency and schedulability. Our evaluation results
show that our generative profiling algorithm
achieves high accuracy while being efficient, and that our co-allocation technique 
substantially improves schedulability compared to a state-of-the-art deadline decomposition method.

\end{abstract}
\section{Introduction}\label{sec:intro}
\noindent
Modern real-time systems are becoming increasingly complex. An automotive application, for instance, typically consists of multiple components that are dependent on one another via input-output data processing flow or synchronization constraints. To meet their increasing resource demands, the applications are commonly deployed on multicore platforms to utilize hardware and software parallelisms. This trend has led to substantial interest in multicore scheduling techniques for the DAG task model, where each task is specified as a directed acyclic graph whose nodes represent subtasks and edges represent precedence constraints between subtasks~\cite{verucchi-rts23-survey}.

A range of DAG scheduling techniques have been proposed, which aim to effectively utilize cores, reduce run-time overheads, and improve schedulability analysis. However, existing solutions focus primarily on the CPUs while ignoring other shared resources, such as the last-level shared cache and the memory bandwidth. This can result in deadline violations, as subtasks running on different cores can interfere with one another by concurrently accessing the shared resources, leading to increased worst-case execution times (WCETs). Recent research has begun to address this issue by developing overhead-aware scheduling and analysis, e.g., to reduce inter-core communication~\cite{shi-rtas24}, to account for memory contention~\cite{casini-rtas20}, or to reduce execution time by cache-aware co-location of  subtasks~\cite{tessler-rtss23}. While these techniques consider shared resources, they do not explicitly {\em control} the resources allocated to subtasks and thus still suffer from interferences caused by concurrent accesses. 

In this work, we take a different road: instead of being simply resource-aware, we propose a {\em co-design}  approach that allocates resources and schedules subtasks in a tightly coupled fashion. By explicitly allocating a 
(dedicated) amount of resources to each concurrent subtask to best meet its needs, we can eliminate the potential interferences due to concurrent accesses while maximizing resource utilization. Our approach uses deadline decomposition to transform each DAG task into an equivalent set of subtasks with release offsets and subtask deadlines, which are scheduled globally on the cores. Unlike state-of-the-art deadline decomposition techniques, however, we co-compute the deadlines of subtasks and their resource budgets together to effectively utilize the available resources, optimize execution progress, and maximize schedulability. 

One challenge towards our co-design approach is the cyclic dependency between deadlines and resource allocation. The deadline of a subtask is generally assigned based in part on its WCET, but its WCET varies depending on how much resource it is given. Assigning a larger deadline enables a smaller resource budget for the subtask and leaves more resources for other concurrent subtasks; however, this also leads to tighter deadlines for successor subtasks (to meet the DAG's deadline) and thus less time for them to execute, which in turn require more resources. One way to go around this problem is to evenly divide resources among cores; however, this approach can result in poor resource utilization and schedulability, since different subtasks tend to exibit different resource demands. 

The second challenge is to derive the multicore resource profile of a workload, such as the number of cache requests, cache misses and retired instructions in each time unit interval, which is needed to match resource allocation with resource demands. Existing work has shown that, even under a fixed resource allocation budget, the resource profile of a workload can change substantially through different phases of an execution~\cite{dna-rtas21}. As it is challenging, if not impossible, to formally analyze such a fine-grained resource profile, the standard approach is to use measurements. However, building a complete resource profile of each subtask under each possible resource allocation budget based on pure measurements is extremely tedious and time consuming. 

\begin{figure*}[t!]
\centering
    \includegraphics[width=.99\linewidth]{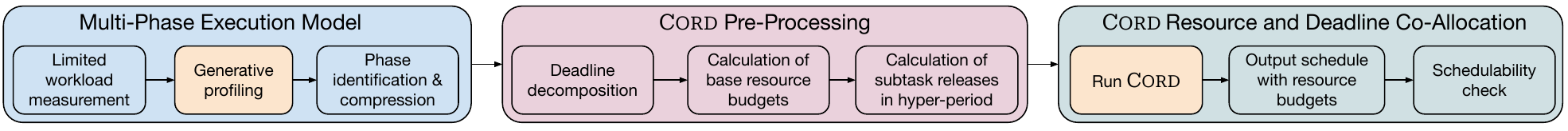}
    \vspace{-1mm}
    \caption{Workflow of our generative profiling and schedule-resource co-design algorithms.} 
    \label{fig:overview}\vspace{-2ex}
\end{figure*}

To bridge this gap, we propose a {\em generative stochastic model} for generating the multicore resource profile of real-time workloads based on limited training data (e.g., obtained by measurements). Our model represents the multicore resource characteristics of a workload at a given time $t$ in an execution as a multi-dimensional resource state $\xi(t)$, where each dimension specifies a computational resource aspect, such as the number of instructions retired, the number of cache requests, or the number of cache misses, in the current time unit interval. Given a set of sample execution paths as training data, with each path consisting of a number of snapshots of the resource state, our goal is to generate the most likely vectorial sample path on the computational resource state $\xi$ conditional on the allocated resource budget. This problem turns out to be non-trivial: first, it is difficult to model $\xi(t)$ from first principle physics since the correlation structure among its components evolve with time; and second, it is impractical to fit parametric statistical models such as a Markov chain by ad hoc gridding or binning the multi-dimensional space of computational resource state. Both these challenges highlight that the desired learning needs to be nonparametric. Our solution lifts the nonparametric learning problem to the infinite dimensional manifold of joint distributions, and solves the constrained maximum likelihood problem over this manifold, which can be done with high efficiency and accuracy. 

We further introduce \sys, a \underline{\bf co}-design of multicore \underline{\bf r}esource allocation and \underline{\bf d}eadline decomposition for DAG tasks based on the generative profiles. \sys exploits the resource profiles of subtasks to make resource allocation and scheduling decisions that effectively utilize shared resources and maximize execution progress. Starting with an initial minimum allocation budget and corresponding deadline assignment for subtasks, our algorithm iteratively distributes available resources to the subtasks that benefit the most, in the process tightening their deadlines as their execution times decrease (as a result of the additional resources given), which in turn informs the selection of subtasks to run on the cores. The result is a set of subtasks with resource-aware release offsets and deadlines, along with a set of timing segments that specify which subtasks are scheduled on the cores and their allocated resources in each segment. By co-allocating resources and deadlines for subtasks, and by exploiting the subtasks' resource profiles in the process, \sys can effectively utilize shared resources to improve execution times and maximize schedulability.
Fig.~\ref{fig:overview} shows the workflow for our generative profiling and co-allocation of multicore resources and deadlines.

In summary, we make the following contributions:
\begin{packeditemize}
\item an efficient method for generative profiling of conditional multi-dimensional resource characteristics based on a multi-marginal Schrödinger bridge solution;
\item a co-design algorithm for multicore resource allocation and deadline decomposition that leverages resource profiles to optimize execution time and schedulability;
\item implementation and evaluation on real benchmarks.  
\end{packeditemize}        
To our best knowledge, this work presents not only the first co-design of multicore resource allocation and deadline decomposition but also the first generative fine-grained resource profiling solution for real-time workloads for DAG scheduling.

\vspace{-1mm}
\section{Related Work}\label{sec:related}
\noindent{\bf CPU scheduling of DAG-based tasks.} 
There exists a large body of work on scheduling and analysis of DAG-based tasks on multiprocessors (see, e.g., the survey in~\cite{verucchi-rts23-survey} and references therein). 
Prior work in this area often focuses on two key directions: (1) 
schedulability and response time analysis methods for a given scheduling algorithm (e.g., global and partitioned EDF with various degrees of preemptions) to improve resource augmentation bounds~\cite{originalspeedup, improvedspeedup, stretch} and tighten worst-case response time~\cite{newrtabound}; and  
(2) parallel scheduling algorithms that aim to effectively utilize cores, reduce run-
time overheads, and improve schedulability (e.g.,~\cite{singlefp1, singlefp2, multifp, pwdm}. Many techniques use deadline decomposition and static priority assignment computed offline to maximize parallelization. For instance, Zhao et al.~\cite{pwdm} models DAG tasks as workload distributions and 
computes an offline priority and core assignment that accumulates the workload distributions to avoid inter-DAG interference. Sun et al.~\cite{edgegen} relies on deep learning to statically generate edges between subtasks to compress the width of the DAG to fit the number of cores. The majority of existing work, however, focuses on only the CPUs while ignoring other shared resources.

\noindent{\bf Resource-aware analysis and scheduling.} 
Recent solutions have started to consider resource interferences in DAG scheduling and analysis. For example,~\cite{shi-rtas24} proposes a scheduling technique that combines subtasks into  execution groups to reduce inter-core communication~\cite{shi-rtas24}. Casini et al.~\cite{casini-rtas20} incorporates the potential overhead due to memory contention in the schedulability analysis. The work in~\cite{tessler-rtss23} co-locates tasks on the same core to reduce cache overhead and improve run-time performance. While these techniques do consider shared resources, unlike our work, they do not explicitly {\em control} the resources allocated to subtasks.

\noindent{\bf Multicore resource allocation.} 
Several multicore resource allocation techniques have been developed in recent years. For example~\cite{xu-rtas19,xu2019holistic} propose holistic resource allocation techniques that find the assignments of tasks, cache and memory bandwidth to cores. Closely related to our work, DNA/DADNA~\cite{dna-rtas21} also exploits the workload characteristics to dynamically adapt resources allocated to a task during an execution. These techniques, however, assume independent tasks whose deadlines are given a priori. We are not aware of any prior work that considers the co-allocation of resources and deadlines to DAG-based tasks.

\noindent{\bf Learning-based prediction in scheduling.}
Learning has been commonly used for predicting workload behaviors and performance in data centers. In the context of real-time systems, there has been a lot of interest in using learning techniques to assist scheduling. For example, prior work has incorporated job size prediction into scheduling of tasks~\cite{predict}. Argawal et al.~\cite{agrawal-ecrts23} shows 
that prediction of run-time behaviors can be used to improve energy efficiency and performance. 
Current techniques are often restricted to the prediction of course-grained information 
such as job size or job response time. In contrast, our work considers much finer-grained resource use characteristics and the time-varying stochastic relationship between multidimensional resource usage \emph{and} a task's available resources. Previous work \cite{bondar-2024-psmsbp,bondar-2024-stochastic} in this area has been concerned with the former, but has not considered the relationship between a software's stochastic time-varying resource usage and the resources allocated to the software.
To our best knowledge, our work is the first to holistically combine generative resource profiling with the co-allocation of multicore shared resources and deadlines. 

\section{System model}\label{sec:model}
\noindent{\bf Multi-DAG tasks.}
The system consists of $n$ DAG tasks, $T = \{\tau_1, \tau_2, \dots, \tau_n\}$, scheduled on a multicore platform. Each task is modeled as a directed acyclic graph whose nodes represent subtasks and edges represent precedence constraints. Each task $\tau_i$ has a period $P_i$ and a relative deadline $D_i$. 
Tasks are released periodically according to their periods. We assume all tasks release their first instance synchronously at time $t = 0$. A source subtask (without a predecessor) is released when its task is released, whereas a non-source subtask is released when all its predecessor subtasks have completed. A task $\tau_i$ is  schedulable if all its subtasks complete their executions within $D_i$ time units from the task's release time.
Each subtask executes a sequential workload (e.g., a function or program), running on at most one core at a time.

\noindent{\bf Platform.}
The platform contains $m$ identical cores that share a set of 
different resource types. For concreteness, we focus on two types of shared resources:  
the (last-level) shared cache and the memory bandwidth. However, 
our proposed method generalizes to other types of shared resources that can be partitioned. We assume the shared cache is partitioned into 
$N_{\mathsf{ca}}$ equal partitions, and the memory bandwidth is partitioned into $N_{\mathsf{bw}}$ equal bandwidth partitions via mechanisms such as Intel's CAT~\cite{intelCAT} for cache and 
MemGuard~\cite{Yun16-memguard-journal} for bandwidth. \\[-2ex]

\noindent{\bf Resource-aware multi-phase execution model for subtasks.} 
Our goal is to co-assign a release offset, a deadline, and a resource budget to each subtask that is released within a hyper-period, to ensure all tasks meet their deadlines. Towards this, we exploit the workloads' resource characteristics to determine an assignment that effectively utilizes resources. We observe that not only is the WCET of a subtask highly dependent on its allocated resource budget, but its execution rate (number of instructions completed  per time unit) under a given budget also changes as the subtask executes through different phases of its program. In addition, the improvement in the subtask's execution rate when given an additional resource budget also varies depending on where it is in its program. 

Motivated by these observations, we propose a resource-aware multi-phase execution model for subtasks to better capture their execution characteristics. We use $\beta$ to denote an allocated resource budget, which is a vector of the number of partitions for each resource type (e.g.,  $\beta = (2, 5)$ represents a resource budget with 2 cache partitions and 5 bandwidth partitions). 
Each subtask $\tau$'s execution is modeled by  
\[
\Theta_{\tau \mid \beta} = (\theta_1, \theta_2, \dots, \theta_k),
\]
which is conditional on an allocated budget $\beta$, where $k$ is the number of consecutive execution phases that capture non-negligible changes in resource characteristics. Each phase $\theta_j = [\theta_j^s, \theta_j^e, \theta_j^r, \theta_j^{\Delta}]$ specifies a start instruction $\theta_j^s$, an end instruction $\theta_j^e$, and a worst-case instruction retirement rate $\theta_j^r$ in this phase under the allocated budget $\beta$.  The WCET of a subtask $\tau$ under an allocated budget $\mathbf{\beta}$ is denoted as $e_{\tau \mid \beta}$.

Intuitively, if $\tau$ is currently at some instruction within the range $[\theta_j^s, \theta_j^e)$ (i.e., in phase $\theta_j$) and is allocated $\beta$ resource budget, it will execute at least $\theta_j^r$ instructions per unit time interval until reaching the ending instruction $\theta_j^e$. 

In addition, $\theta_j^{\Delta}$ captures the improvement in execution rate if $\tau$ is allocated extra resources. Specifically, $\theta_j^{\Delta}[\beta']$ specifies the increase in $\tau$'s execution rate if it is given an extra budget of $\beta'$.
Note that $\theta_k^e$ is $\tau$'s total number of instructions, $\theta_1^s = 0$, and $\theta_j^e = \theta_{j+1}^s$ for all $1\le j < k$.

\noindent{\bf Problem statement.} Given the above setting, our objectives are twofold: 
(1) learn a generative resource profile of a given workload that can be used to construct a multi-phase execution rate model for each subtask; and 2) develop a co-design algorithm that uses this model to compute the release times and deadlines for subtasks, together with the resource budget allocated to each subtask at each scheduling decision point (i.e., a release time or a deadline of a subtask). Our overarching goal is to effectively utilize shared resources, maximize execution progress, and increase schedulability.

\section{Generative multicore resource profile for real-time workloads}\label{sec:prediction}
\noindent
In this section, we present a method for generating the multicore resource profile of a workload, which can be used to construct the multi-phase execution model for subtasks. 

The resource profile of a workload is represented as $d$-dimentional resource state  
\begin{align}
\xi = (\xi_1, \xi_2, \dots, \xi_d)^{\top}\in\mathcal{X}\subset\mathbb{R}^d,
\label{defxi}    
\end{align}
conditional on an allocated budget 
\begin{align}
\beta = \left(\beta_1,\beta_2, \hdots, \beta_b\right)^{\top}\in\mathcal{B}\subset\mathbb{R}^b.
\label{defbeta}    
\end{align} 
Each component of $\xi$ captures a specific resource aspect of the workload at a given time when it is executed under the allocated budget $\beta$. An example of $d = 3$ resource aspects are the number of retired instructions ($\xi_1$), the number of last-level cache requests ($\xi_2$), and the number of cache misses ($\xi_3$). When $b=2$, components of the resource budget $\beta=(\beta_1,\beta_2)^{\top}\in\mathcal{B}\subset\mathbb{R}^{2}$ can, for instance, denote all possible combinations of the allocated cache $\beta_1$, and allocated memory bandwidth $\beta_2$.

Given resource budget $\beta$, the resource state $\xi$ varies with time $t$, and the different resource aspects at a given time are correlated. \emph{In other words, $\xi(t)$ is a stochastic process conditioned on an allocated budget $\beta$}. We denote this conditional stochastic process as 
\begin{align}
\xi(t)\mid\beta.
\label{DefConditionalProcess}    
\end{align}
{The stochastic process \eqref{DefConditionalProcess} is correlated both spatially (across different components of $\xi$ at the same time) and temporally (across $\xi$ vectors at different times). Even when $\beta$ and the initial condition $\xi(t=0)$ are fixed, repeated executions of the same software on the same hardware result in different sample paths of $\xi(t)$. In Sec. \ref{sec:generative}, we will outline the learning and generative sampling from the most likely joint statistics of the conditional stochastic process $\xi(t)\mid\beta$.}

\subsection{Measurements} \label{sec:measurement}
\noindent One way to obtain the resource profile of a workload is by extensive measurements. For instance, we can assign a fixed resource budget $\beta$ to the workload, pin it to a dedicated core, then execute the workload over many trials. For each trial, we can use the CPU's performance counters to periodically collect resource aspects such as the total number of retired instructions $(\xi_1)$, the number of cache requests $(\xi_2)$, and the number of cache misses $(\xi_3)$. This provides a set of snapshots of the state at measured time points for each sample run. By repeating the process for all possible allocated budgets $\beta$, we can obtain a collection of state snapshots at different time points for a finite set of runs, conditioned on $\beta$. 

However, obtaining a fine-grained resource profile from direct measurement is practically infeasible. For example, with $N_{\mathsf{ca}} = 20$ cache partitions and $N_{\mathsf{bw}} = 20$ bandwidth partitions, there are $N_{\mathsf{ca}}\times N_{\mathsf{bw}}= 400$ possible allocated budget $\beta$ for each subtask. For each $\beta$, we require many runs per subtask and a large number of snapshots per run. 
Because of the overheads, it is not possible to obtain a temporally fine-grained profile that contains snapshots at small time increments or at any arbitrary time $t$. 
To circumvent this problem, we propose to use a small subset of measurement snapshots only as training data, and present an efficient algorithm for learning the most likely conditional resource profile for the workload. 

\subsection{Generative resource profiling algorithm}\label{sec:generative}
\noindent 
Recall that the set $\mathcal{B}$ in \eqref{defbeta} denotes the set of all possible resource allocations to a software. For instance, when $b=2$ with $N_{\mathsf{ca}}$ partitions of {allocated} cache and $N_{\mathsf{bw}}$ partitions of {allocated} memory bandwidth, then {$\mathcal{B}\subset\mathbb{R}^{2}$ is a finite set of cardinality $N_{\mathsf{ca}}\cdot N_{\mathsf{bw}}$.} 
For 
{any}
$\beta\in\mathcal{B}$ we desire a profile $\xi(t) \mid \beta$, but if $\mathcal{B}$ is large, it may be feasible to obtain these profiles empirically for only a subset $\mathcal{B}'\subsetneq\mathcal{B}$. Moreover, technical limitations in the profiling mechanism or the software itself may have these `profiles' consist of a temporally coarse set of snapshots, which are empirical distributions of the conditional random vector
\begin{align}
\xi(t_\sigma)\mid\beta
\label{defSnapshotDistribution}  
\end{align} 
for measurement instances $t_{\sigma\in\llbracket n_s\rrbracket}$ and $\beta\in\mathcal{B}'$. Here and hereafter, we adopt the finite set notation
$$\llbracket n_s\rrbracket := \{1,2,\hdots,n_s\}.$$
In other words, \eqref{defSnapshotDistribution} is the usage of computational resources $\xi$ at time $t_\sigma$ conditional on the allocated budget $\beta$.

For a \emph{fixed} resource allocation $\beta$, recent work \cite{bondar-2024-psmsbp,bondar-2024-stochastic} computed the most likely distributional path $\mu_t \mid \beta$ for $t\in[t_1,t_{n_s}]$ from snapshots $\mu_{\sigma\in\llbracket n_s\rrbracket}\mid\beta$. In this work, we extend these methodologies to generate distributions (and thereby, profiles) for resource allocations wherefor data is not available.

Similar to \cite{bondar-2024-psmsbp,bondar-2024-stochastic}, we compute the \emph{multi-marginal Schr\"odinger bridge (MSB)} to find the \emph{most likely measure-valued path} between a set of resource usage probability distributions. Specifically, given a set of resource allocations $\{\beta^j\}_{j\in\llbracket n_b\rrbracket}$ where for each we have the corresponding $n_d$ profiles $\{\xi^{i,j}(t)\}_{i\in\llbracket n_d\rrbracket}$, we construct the \emph{empirical distributions}
\begin{align}
\mu_\sigma := \frac{1}{n_dn_b}\sum_{i=1}^{n_d}\sum_{j=1}^{n_b}\delta(\eta-\eta^{i,j}(t_\sigma)), \quad \forall\sigma\in\llbracket n_{s}\rrbracket,
\label{defmusigmaempirical}    
\end{align}
supported over the \emph{augmented state} $ \eta:=\begin{bmatrix}\xi & \beta\end{bmatrix}^\top \in \mathcal{X}\times\mathcal{B}\subset\mathbb{R}^{d+b}$, where $\delta$ denotes the Dirac delta. The $n_{s}$ measurement instances $t_{\sigma\in\llbracket n_s\rrbracket}$ are the times
$$ 0=t_1<t_2<\dots<t_{n_s-1}<t_{n_s},$$
at which `snapshots' of $\xi$ were taken for each of the $n_d$ profiles. The MSB between the $n_s$ empirical distributions, thus constructed, finds the most likely measure-valued path $\mu_t$, where $\eta(t)\sim\mu_t$ $\forall t\in[t_1,t_{n_s}]$,
satisfying the distributional constraints 
\begin{align}
\eta(t_\sigma)\sim\mu_\sigma \qquad\forall\sigma\in\llbracket n_s\rrbracket.
\label{DistributionalConstraints}
\end{align}
We then apply Bayes' theorem to obtain the \emph{most likely conditional joint distribution of resource usage} as
\begin{align}
\xi(t) \mid \beta \sim \frac{\mu_t}{\int_\mathcal{X} \mu_td\xi} \:.
\label{ComputeConditional}    
\end{align}
The numerator in \eqref{ComputeConditional} is the joint distribution over the augmented state $\eta(t)$ computed from the MSB, and the denominator is the marginal distribution of the resource budget $\beta\in\mathcal{B}\subset\mathbb{R}^{b}$. As the denominator is independent of time $t$, it can be pre-computed before solving the MSBP.  

Details on computing the MSB itself -- i.e. solving the MSB Problem (MSBP) -- between known resource usage probability distributions indexed by temporal measurement instances $t_{\sigma\in\llbracket n_s\rrbracket}$ can be found in the Supplementary Materials. 

By re-sampling the computed conditional distributions \eqref{ComputeConditional} we are able to generate high conditional probability samples, and thus generative or synthetic profiles. This re-sampling can, for instance, be done by simply returning the top few high probability samples since the distribution-values evaluated at the samples, and not just the samples of \eqref{ComputeConditional}, are available from MSB computation. Alternatively, this re-sampling can be done using existing diffusion models \cite{sohl2015deep,song2019generative,song2020score} for generative AI. 

We then use these synthetic profiles to predict a software's most-likely resource usage at any time $t\in[t_1,t_{n_s}]$ subject to any given resource allocation $\beta\in\mathcal{B}$. In previous work \cite{bondar-2024-psmsbp,bondar-2024-stochastic}, this was done only under a fixed $\beta$, i.e., where $\eta=\xi$.

\subsection{Multi-phase execution model from generative profile}\label{sec:phases}

\noindent For each subtask $\tau$, and each resource budget $\beta$, our generative profile provides the number of retired instructions ($\xi_1$), the number of last-level cache requests ($\xi_2$), and the number of cache misses ($\xi_3$) at each time point in the subtask's execution.
Our goal is to construct a series of execution phases $\Theta_{\tau \mid \beta} = (\theta_1, \theta_2, \dots, \theta_k)$ for each $\tau$ under each resource budget $\beta$ by identifying consecutive segments of its generated profile that display similar resource characteristics.

A natural approach for identifying similar groups in our three-dimensional profile (instructions, cache requests, and cache misses) is to use a $k$-means clustering technique. 
In our case, we use a Gaussian Mixture Model (GMM)~\cite{reynolds2009gaussian} to identify clusters in the combined profiles of each subtask, although this can be accomplished using various other methods. 
Since the expected number of execution phases, $k$, is not known in advance, we test $3 \leq k \leq 20$ and use the Davies-Bouldin index~\cite{davies-pami79} to assign a quality score to a particular clustering. 
We then select the smallest $k$ value that demonstrates non-diminshing returns in this score.

The GMM assigns each time point in the profile a cluster value between 1 and $k$.
We then compress the 
result by combining consecutive time points with the same phase number. 
The result is a series of phases $\Theta_{\tau \mid \beta} = (\theta_1, \theta_2, \dots, \theta_k)$, where each phase $\theta_j$ has a start instruction $\theta_j^s$, an end instruction $\theta_j^e$, and a worst-case execution rate $\theta_j^r$ that is calculated by taking the minimum execution rate of any time point in cluster $j$.

The final parameter of the model is the table $\theta_j^{\Delta}$. 
Recall that if $\tau$ is in phase $\theta_j$ (based on its current instruction count), 
then $\theta_j^{\Delta}[\beta']$ represents the expected increase in execution rate of $\tau$ if we give it $\beta'$ additional resource budget.
A simple way to calculate $\theta_j^{\Delta}[\beta']$ is to find $\theta_i \in \Theta_{\tau \mid \beta + \beta'}$, that is, the phase of $\tau$ under the new budget $\beta + \beta'$. 
One could then calculate the difference between the current execution rate $\theta_j^r$ and the new rate $\theta_i^r$ to determine $\theta_j^{\Delta}[\beta']$.
However, we also want to account for how much of our resource budget is still available.

For example, suppose that giving $\tau$ an additional partition of cache does not cause its execution rate to increase, but the rate increases by a factor of 10 when we give it 3 partitions.
If we have 3 remaining cache partitions, but allocate resources one by one without incorporating this knowledge, we might distribute these partitions to less resource-sensitive subtasks, missing opportunities to maximize resource efficiency.
Therefore, to calculate $\theta_j^{\Delta}[\beta']$ we take the average rate change $\theta_i^r - \theta_j^r$ for all combinations of additional resources that could be given out in the future.
For simplicity, we call $\theta_j^{\Delta}[\beta']$ the expected improvement in execution rate given additional resource $\beta'$; however, its value is calculated by looking ahead at the changes in rates that could be achieved given the remaining resource budget. 

The constructed multi-phase execution model provides an efficient way to look up a subtask's current execution rate and expected change in execution rate when given more resources. We next present an algorithm that utilizes this information to effectively allocate resources and deadlines to subtasks. 

\section{Resource-allocation and deadline decomposition co-design}\label{sec:algorithm}
\noindent
We now present \sys, an algorithm for co-allocating resources and deadlines for subtasks  
that exploits the multi-phase execution $\Theta$ of subtasks
to maximize resource utilization, reduce execution time, and optimize schedulability.

\subsection{Overview}\label{algo:overview}
\noindent At a high level, \sys works by considering the subtasks of all 
tasks in the system that are released within a hyper-period. 
Starting with an initial release time, deadline, and base (`minimum') resource budget 
$\beta^{init}_{\tau}$ for each subtask $\tau$, \sys uses an iterative algorithm 
to compute new resource budgets and deadlines for subtasks. 
The release times and completion times of the subtasks form a set of \emph{decision points}, 
at which we make scheduling and resource allocation decisions.



At each decision point (from time 0 onwards), \sys computes a new resource budget 
for each subtask that is currently ready. 
The goal is to maximally reduce WCETs by redistributing resources to the subtasks 
that would benefit the most from extra resources, while respecting subtasks' 
deadlines (by ensuring they are allocated sufficient resources to complete before 
their deadlines, if at all possible). For this, \sys utilizes the multi-phase 
execution rate $\Theta_{\tau\mid \beta}$ of each subtask $\tau$ under an allocated 
budget $\beta$ (constructed in Section~\ref{sec:prediction}) to determine the 
subtasks that would benefit the most (i.e., the ones with the highest increase in 
execution rates) with the extra resources. As more resources are allocated to 
these subtasks, their WCETs shrink and we adjust their deadlines accordingly. 
Under the global EDF scheduling policy, we can directly determine which subtasks 
should be scheduled on the cores based on these deadlines. 
As the deadlines of some subtasks are shortened, the set of subtasks with the earliest deadlines may change, causing the set of tasks that are chosen by EDF to change. 
Our algorithm iteratively recomputes the resource budgets for subtasks
until it is no longer possible to shrink their WCETs given the available 
resources. When this happens, it will move to the next decision point.

The output of our algorithm is a static schedule 
for one hyper-period, which is made of a series of consecutive segments, with the exact ($\le m$) subtasks 
that will be executed on the $m$ cores and the allocated resource budgets for 
the scheduled subtasks during each segment. Note that a subtask execution may span multiple consecutive 
or non-consecutive segments, and its allocated resource budget may also change across these segments.
Since the tasks are assumed to release their first job synchronously, the system is schedulable if 
the static schedule produced by our algorithm ensures that each subtask’s effective WCET 
under the allocated budget plus its release time does not exceed the overall DAG task's deadline.

Before discussing our algorithm in detail, let us define some notation.
Let $H$ be the hyper-period of the tasks in the system, i.e., the least common multiple of the tasks' periods. 
Let $K_i$ be  the number of releases of $\tau_i$ in one hyper-period (i.e., $K_i = H/P_i$ where $P_i$ is $\tau_i$'s period). 
Then, we have a series of fixed release points $A_i = \{(k-1)\cdot P_i\; | \;1 \leq k \leq K_i\}$ 
for the jobs of $\tau_i$ in one hyper-period. We call $A_i$ the \emph{anchor points} of $\tau_i$. 

\subsection{Computing base resource budgets and initial deadlines}\label{sec:init}
\noindent We begin with the computation of the base resource budgets for subtasks, together with their release times and deadlines.
We propose two approaches: a {\em greedy} approach that optimizes   
resource efficiency, and a {\em deadline-aware} approach that takes into account tasks' deadlines.
We describe them below.

{\bf Greedy approach:}
In this approach, we give each subtask $\tau$ the {\em absolute minimum} resource budget 
$\beta^{init}_{\tau} = \mathbf{1}$ (i.e., one cache partition and one bandwidth partition). 
We then assign a release time and deadline to $\tau$ based on its WCET under this minimum budget, 
denoted as $e_{\tau \mid \mathbf{1}}$. Specifically, the deadline of $\tau$ is set equal to 
its release time plus $e_{\tau \mid \mathbf{1}}$. Its release time is set to the latest 
deadline of any of its predecessors, or the anchor point of its task instance if $\tau$ 
has no predecessors. By using the minimum resources initially for the base allocation, 
this approach leaves the maximum ``unused'' resources for our co-allocation algorithm to work with,
enabling more opportunities for the algorithm to optimize resource use efficiency.

{\bf Deadline-aware approach:}
The greedy approach does not consider tasks' deadlines and thus may impact those with 
tight deadlines (which generally require more resources than the absolute minimum to finish 
before their deadlines). Our deadline-aware approach avoids this by assigning the 
minimum resource budget that each subtask would need for the overall DAG task to 
complete by its deadline, assuming sufficient cores. Towards this, we first apply an existing deadline decomposition 
method to assign release times and deadlines to subtasks, where the WCET of a subtask  
is assumed to be its WCET with the maximum amount of resources. By iteratively taking away resource partitions, we then determine the {\em minimum}  
resource budget $\beta^{init}_{\tau}$ that each subtask $\tau$ would need to 
ensure its resulting WCET plus its release time does not exceed 
its deadline. For our evaluation, we adopted the deadline decomposition method from~\cite{stretch} to derive the release times and deadlines for subtasks; however,
any other deadline decomposition method that maximizes subtask slack can be used here. 

The computed subtasks' release times, deadlines, and the base budget allocation $\beta^{init}$
(using either approach) will serve as inputs to our 
co-allocation algorithm, which we detail next.

\subsection{Resource-deadline co-allocation algorithm}\label{algo:iterative}
\noindent 
Algorithm~\ref{alg:cord} shows the pseudocode of our resource-deadline 
co-allocation algorithm (\algo). 
Using the result from Section~\ref{sec:init} as the initial allocation,
it computes a static schedule, $\Sched$. Each element $\Sched[i]$ is 
a tuple containing a decision time point $t_i$, 
the set of (at most $m$) subtasks to execute at $t_i$, 
and the assigned resource budgets for these subtasks 
(where $0 = t_0 < t_1 < \cdots < t_k < H$, and $H$ is the hyper-period).

The algorithm takes as input the following parameters: 
the set of all subtask releases $J$ in the hyper-period; 
the set $A$ that contains the anchor points (fixed release times) of 
the tasks' instances in the hyper-period; and 
the base resource budget $\beta_{\tau}^{init}$, the release time $r_{\tau}$, 
and the deadline $d_{\tau}$ for all $\tau \in J$. 
For its co-allocation, \algo also takes as input 
the total number of instructions $\mathsf{maxIns}_{\tau}$ and 
the multi-phase execution model $\Theta_{\tau}$ for each subtask $\tau$ 
(constructed in Section~\ref{sec:phases}). We use $R_{\max}$ to denote the vector 
of maximum number of partitions per resource type provided by the platform 
(e.g., $R_{\max} = (N_{\mathsf{ca}}, N_{\mathsf{bw}})$ on a platform with 
$N_{\mathsf{ca}}$ cache partitions and $N_{\mathsf{bw}}$  bandwidth partitions).
Throughout the algorithm, $t$ denotes the current decision point at
which we compute the schedule for, $\tnext$ denotes the next decision point, 
$B$ denotes the future anchor points, $\Retired_{\tau}$  
denotes the number of instructions $\tau$ has already completed (retired) at $t$,  
$Q$ denotes the set of ready subtasks at $t$, and 
$S$ and $R_s$ denote the set of scheduled subtasks (to execute on the cores at $t$) 
and the total resource budget assigned to these subtasks, respectively. 
Finally, $\beta$ denotes the current resource allocation (initialized to $\beta^{init}$),
and $r_{\tau}$, $c_{\tau}$ and $d_{\tau}$ denote the release time, completion time and deadline of $\tau$ under the current budget $\beta_{\tau}$, respectively.

The algorithm begins by initializing the variables (Lines~\ref{line:init-begin}--\ref{line:init-end}). 
It sets $\Sched$ to be empty and its current index $i$ to zero.
For each subtask $\tau \in J$, it initializes the retired instruction count $\Retired_{\tau}$ to 0 and
the  flag $\done_{\tau}$ to false to indicate $\tau$ has not completed its execution. It also calculates the completion time $c_{\tau}$ of each $\tau \in J$ under the base resource budget.  
It then sets $t = 0$ as the current decision point for re-allocation   
and set the next decision point $\tnext$ as the earliest release time or completion time of any of the jobs in $J$ that occurs after $t$. 
Next, \algo initializes the set of future anchor points $B$ and 
constructs the set of ready subtasks $Q$ to consider 
for scheduling at time $t$. After this initialization, it proceeds to the main while-loop 
(Lines~\ref{line:8}--\ref{line:endwhile}). In each iteration, \algo performs resource-deadline
co-allocation in three main steps:

{\bf Step 1 (Lines~\ref{line:9}--\ref{line:15}):} For $\tau \in Q$, \algo first initializes the current budget $\beta_{\tau}$ 
to the base budget and saves the initial deadline (Line~\ref{line:beta-init}). It then selects the $m$ subtasks (if any) with the smallest
deadlines in $Q$ under their current budgets to be the scheduled set $S$, and computes the total budget $R_s$
of these subtasks (i.e., $R_s = \sum_{\tau \in S} \beta_{\tau}$) (Line~\ref{line:10}). 

If $R_s$ is more than $R_{\max}$ (the maximum resources provided by the platform), \algo will iteratively take resources away from the subtask with the most slack in $S$ until the total budget $R_s$ used by the scheduled set $S$ is within $R_{\max}$ (Lines~\ref{line:11}--\ref{line:15}). 



\begin{algorithm}[t!]\raggedright
    \caption{Resource-Deadline Co-allocation Algorithm}
    \label{alg:cord}
    \small
    \begin{algorithmic}[1]

        \State \textbf{Input: } $J$, $A$, $\beta^{init}$, and 
                $\{ r_{\tau},\, d_{\tau},\, \mathsf{maxIns}_{\tau},\, \theta_{\tau}\, \mid\, \tau \in J \}$.
        \State \textbf{Output: } $\Sched$ \Comment{Computed schedule and allocated budgets}
 
        \State $\Sched = []$; $i = 0$;  \Comment{Array and index for output schedule} \label{line:init-begin}

        \ForAll{$\tau \in J$} 
            \State $\Retired_{\tau} \leftarrow 0$; $\done_{\tau} \leftarrow \textbf{false}$; $c_\tau \leftarrow r_\tau + e_{\tau \mid \beta^{init}}$;
        \EndFor
        \State $t \leftarrow 0$;   \Comment{Current decision point} \label{line:t0}
        \State $\tnext \leftarrow \min \{ r_{\tau} > 0, c_{\tau} > 0 \,\mid\, \tau \in J \})$  \Comment{Next decision point}  \label{line:tnext-init}  
        \State $B \leftarrow A \setminus \{ 0 \}$ \Comment{Set of future anchor points} \label{line:B0}
        \State $Q \leftarrow \{ \tau \in J \mid r_{\tau} = 0 \}$ \Comment{Ready subtasks} \label{line:init-end}

        \While{\textbf{true}} \label{line:8}
            \ForAll{$\tau \in Q$}:  \label{line:9}
                $ d^{init}_\tau \leftarrow d_\tau$;
                $\beta_{\tau} \leftarrow \beta^{init}_{\tau}$ \Comment{Assign base} \label{line:beta-init}
            \EndFor

            \State $(S, R_s) \leftarrow \getSched(Q)$ \Comment{$R_s$: budget of $S$} \label{line:10}
            \While {$(R_s > R_{\max})$}  \label{line:11}
                \State $(\tau, \delta_{\tau})  \leftarrow \maxSlackTask(\beta, S, R_s)$ \label{line:12}
                \State $\beta_{\tau} \leftarrow \beta_{\tau} - \delta_{\tau}$    \Comment{Reduce budget of max slack subtask} \label{line:13}
                \State $R_s \leftarrow R_s - \delta_{\tau}$  \Comment{Update the total budget used by $S$}
            \EndWhile \label{line:15}

            \While{\textbf{true}} \label{line:16}
                \State $(\tau, \delta_{\tau}) \leftarrow \resourceAlloc(Q, S, R_s, \beta, \Retired, t, \tnext)$ \label{line:17}
                \If{$\tau = $ \textbf{null}} \textbf{break} \Comment{Cannot give more resources}\label{line:18}
                \EndIf

                \State $\beta_{\tau} \leftarrow \beta_{\tau} + \delta_{\tau}$ \Comment{Give extra resource to $\tau$} \label{line:19}
                \State $d_{\tau} \leftarrow d_\tau - (c_{\tau} - \finishTime(\tau, \beta_\tau, \Retired_{\tau}, t, \tnext))$ \label{line:20}
                \State $c_{\tau} \leftarrow \finishTime(\tau, \beta_\tau, \Retired_{\tau}, t, \tnext,)$ \label{line:19b}

                \If{ $\tau \notin S$} \label{line:21}
                    \State $\tau^{s} \leftarrow \argmax_{\tau' \in S}\;\; d_{\tau'}$  \Comment{Max deadline subtask in $S$}
                    \If{ ($d_{\tau} < d_{\tau^s}$ \textbf{and} $R_s - \beta_{\tau^s} + \beta_{\tau} \leq R_{\max}$)} 
                        \State $S \leftarrow (S \setminus \{ \tau^s\}) \,\cup\, \{ \tau \} $ \Comment{Swap $\tau_s$ with $\tau$}
                    \EndIf
                \EndIf \label{line:24}

                \If {$c_{\tau} \ge \tnext $} \label{line:25}
                    \State $(S, R_s) \leftarrow \getSched(Q)$ \label{line:26}
                \Else
                    \State $(S, R_s) \leftarrow \resetSegment(\tau, Q, \Retired,  \beta^{init}, t)$\label{line:28}
                    \State $\tnext \leftarrow c_{\tau}$ \label{line:tnext-reset}
                \EndIf
            \EndWhile \label{line:28}

            \ForAll{$\tau \in Q \setminus S$} \Comment{$\tau$ is an unscheduled subtask} \label{line:29}
                \State $\beta_{\tau} \leftarrow \beta^{init}_\tau$; $d_\tau \leftarrow d_\tau^{init}$; \Comment{Reset to base} \label{line:30}
                \State $c_{\tau} \leftarrow \finishTime(\tau, \beta_\tau, \Retired_{\tau}, \tnext, \infty)$ \label{line:31}
            \EndFor

            \ForAll{$\tau \in S$} \Comment{Update scheduled subtasks \& successors} \label{line:32}
                \State $\Retired_{\tau} \leftarrow \computeRetiredInstruction(\tau,  \beta_\tau, \Retired_{\tau}, t, \tnext)$ \label{line:33}
                \If{$\Retired_{\tau} = \mathsf{maxIns}_{\tau}$}  \Comment{$\tau$ finished}
                    \State $c_{\tau} \leftarrow \tnext$;\; $\done_{\tau} \leftarrow \textbf{true}$ \label{line:34}
                    \State $\mathsf{readySucc} \leftarrow \shiftSuccessors(\tau)$  
                    \label{line:35}
                    \State $Q \leftarrow Q \setminus \{\tau\} \cup \mathsf{readySucc}$
                    \State 
                \EndIf
            \EndFor
               
            \State $\Sched[i\hbox{++}] \leftarrow ( t, \;\{ (\tau, \beta_{\tau}) \mid \tau \in S \} )$ 
                \Comment{Save allocation} \label{line:36}
            \If{$Q = \emptyset \,\land\, B = \emptyset$} \textbf{break} 
                        \Comment{Complete algorithm} \label{line:38}
            \EndIf 

            \If{$Q = \emptyset$} $t \leftarrow \min(B)$ \textbf{else} $t \leftarrow \tnext$  
            \label{line:39}
            \EndIf
            
            \ForAll{($\tau \in J \,\land\, r_{\tau} = t \,\land\, \mathsf{parents}(\tau) = \emptyset$)} \label{line:release-anchor}
               \State $Q \leftarrow Q \cup \{ \tau \}$ \Comment{Release ready source subtasks; add to Q}
            \EndFor
            \State $B \leftarrow B \setminus \{ t\}$ \label{line:B-update} \Comment{Update the set of future anchor points}
            
            \State $\tnext = \min\{\;\min(B),\; \min_{\tau \in Q} c_{\tau}\,  \}$ \label{line:tnext-update}\Comment{Next decision point}
        \EndWhile \Comment{Complete on segment} \label{line:endwhile}

        \ForAll{$\tau \in J$}   \label{line:41}
            \State $\tau_{i,k} \leftarrow \mathsf{taskInstance}(\tau)$ 
            \If{$c_{\tau} > (k-1)*P_i + D_i$} \textbf{return} $\textbf{false}$ \Comment{Unschedulable}
            \EndIf
        \EndFor \label{line:43}
        \State \textbf{return} $\mathsf{true}$ \Comment{Schedulable}
        \State \textbf{Output:} $\Sched$ 

    \end{algorithmic}
\end{algorithm}

\setlength{\textfloatsep}{2mm}
\begin{algorithm}[t]\raggedright
    \caption{$\resourceAlloc$}
    \label{alg:resalloc}
    \small
    \begin{algorithmic}[1]
        \State \textbf{Input: }$Q, S, R_s, \beta, \Retired, t, \tnext$
        \State \textbf{Output: }$(\tau^{\best}, \delta_{\tau})$
        \State $\tau^{\best} \leftarrow \textbf{null}$
        \State $\{\beta_1, \beta_2\} \leftarrow R_{\max} - R_s$
        \Comment Check for available resource types
        \ForAll{$\tau \in Q$}
            \If{$\tau \notin S$ \textbf{and} $\beta_\tau = R_{\max}$} \textbf{continue} \EndIf
            \If {$\tau \in S$ \textbf{}and $\{\beta_1, \beta_2\} = \{0, 0\}$} \textbf{continue}  \EndIf 
            \State $\Retired^{\tnext}_{\tau} \leftarrow \computeRetiredInstruction(\tau,  \beta_\tau, \Retired_{\tau}, t, \tnext)$
            \State $\theta_{i} \leftarrow \Theta_{\tau \mid \beta}[\Retired_\tau]$
            \Comment Find current phase $\theta_i$ using $\Retired_\tau$
            \State $B_\tau' \leftarrow \{0, 0\}$
            \Comment Track preferred resource type
            \State $\theta^{\Delta}_\tau \leftarrow 0$;
            \Comment Weighted sum of all $\theta^{\Delta}$'s in segment
            
            \While{$\theta_i^s < \Retired^{\tnext}_{\tau}$}
                \State $\phaseStartIns \leftarrow \max(\theta^s_{i}, \Retired_\tau)$;  $\phaseEndIns \leftarrow \min(\theta^e_{i}, \Retired^{\tnext}_{\tau})$; 
                \State $\beta_j \leftarrow \argmax_{\beta_j \in \{\beta_1, \beta_2\}}\;\; \theta^{\Delta}_{i}[\beta_j]$
                \State $\theta^{\Delta}_\tau \leftarrow \theta^{\Delta}_\tau + \theta^{\Delta}_{i}[\beta_j] / (\phaseEndIns - \phaseStartIns)$
                \State $\beta_\tau'[j] \leftarrow \beta_\tau'[j] + (\phaseEndIns - \phaseStartIns)$
                \State $i \leftarrow i + 1$
            \EndWhile
        \EndFor
        \State $\tau^{\best}\leftarrow \argmax_{\tau \in Q}\;\; \theta^{\Delta}_\tau$
        \Comment Give resource to $\tau$ with largest $\theta_i^{\Delta}$
        \State $\delta_\tau \leftarrow \argmax_{\beta_j \in \{\beta_1, \beta_2\}}\;\; \beta_{\tau^{\best}}'[j]$ \Comment Resource type that most increased $\theta_i^{\Delta}$
        \State \textbf{Output: }$(\tau^{\best}, \delta_{\tau})$
    \end{algorithmic}
\end{algorithm}

\begin{algorithm}[t!]\raggedright
    \caption{$\finishTime$}
    \label{alg:computed}
    \small
    \begin{algorithmic}[1]
        \State \textbf{Input: }$\tau, \beta_\tau, \Retired_{\tau}, t, \tnext$ \Comment Given that $\beta_\tau$ resets to base at $\tnext$
        \State \textbf{Output: }$c_{\tau}$ \Comment Calculate estimated finish time

        \State $\Retired^{\tnext}_{\tau} \leftarrow \computeRetiredInstruction(\tau,  \beta_\tau, \Retired_{\tau}, t, \tnext)$
            \State $\theta_{i} \leftarrow \Theta_{\tau \mid \beta}[\Retired_\tau]$
            \Comment Find current phase $\theta_i$ using $\Retired_\tau$
            \State $c' \leftarrow 0$; $k \leftarrow |\Theta_{\tau \mid \beta}|$;
            \Comment Get total number of phases in $\Theta_{\tau \mid \beta}$
            \While{$i \leq k$}
                \State $\phaseStartIns \leftarrow \max(\theta^s_{i}, \Retired_\tau)$;  $\phaseEndIns \leftarrow \min(\theta^e_{i}, \Retired^{\tnext}_{\tau})$;
                \State $c' \leftarrow c' + (\phaseEndIns - \phaseStartIns) / \theta^{r}_{i}$ \Comment Get WCET using $\theta^{r}_{i}$
                \If {$\theta_i^e > \Retired^{\tnext}_{\tau}$}
                    \State $\theta_{i} \leftarrow \Theta_{\tau \mid \beta^{init}}[\Retired^{\tnext}_{\tau}]$ \Comment Resource budget resets after $\tnext$ 
                    \State $k \leftarrow |\Theta_{\tau \mid \beta^{init}}|$ \Comment Using new phase model now
                    \State $\Retired^{\tnext}_{\tau} \leftarrow \mathsf{maxIns}_{\tau}$ 
                \Else
                \State $i \leftarrow i + 1$ \Comment Go to next phase in current model
                \EndIf
            \EndWhile
            \State $c_\tau \leftarrow c_\tau + c'$
            \State \textbf{Output: }$c_{\tau}$

    \end{algorithmic}
\end{algorithm}

\begin{algorithm}[t!]\raggedright
    \caption{$\shiftSuccessors$}
    \label{alg:shiftsucc}
    \small
    \begin{algorithmic}[1]
        \State \textbf{Input: }$\tau$
        \State \textbf{Output: }$\mathsf{readySucc}$

        $\mathsf{readySucc} \leftarrow \emptyset$

        \ForAll{$\tau_{succ} \in \mathsf{successors}(\tau)$} 
            \If{$\done_{\tau_{pred}} \;\forall \tau_{pred} \in \mathsf{predecessors}(\tau_{succ})$} 
            \State $\mathsf{readySucc} \leftarrow \mathsf{readySucc} \cup \{\tau_{succ}\}$ \Comment{All parents done}
            \State $r_{\tau_{succ}} \leftarrow c_\tau$
            \EndIf
        \EndFor

        \State \textbf{Output: }$\mathsf{readySucc}$

    \end{algorithmic}
\end{algorithm}

{\bf Step 2 (Lines~\ref{line:16}--\ref{line:28}):} 
In this step, \algo iteratively reallocates resources of the ready subtasks to shrink WCETs and deadlines. 
In each iteration, it first calls the function $\resourceAlloc$ 
 (shown in Algorithm~\ref{alg:resalloc}) 
to determine the subtask $\tau$ in $Q$ that would benefit the most from extra resources during the timing segment from $t$ to the immediate next decision point $t_{\mathsf{next}}$  and the amount of extra resource budget $\delta_{\tau}$ to give to this task (Line~\ref{line:17}).

Intuitively, $\resourceAlloc$ uses the multi-phase execution model $\theta_{\tau}$ to determine which subtask 
$\tau$ would have the maximum increase in execution rate if it were given all the remaining $R_{\max} - R_s$ unused
resource budget. This is done by considering the phases $\theta_j$ that each subtask $\tau$ will execute from its current instruction until reaching the next decision point under its current resource budget. Using the number of instructions that $\tau$ will execute within these phases as the weights, we then compute a weighted sum of the execution rate increments $\theta_j^{\Delta}[R_{\max} - R_s]$ across these phases if $\tau$ were given all unused resources.  The subtask with the highest weighted sum is the one that would benefit the most from the extra resources, and thus will be selected.  
Once the subtask is selected, $\resourceAlloc$ determines the extra budget $\delta_{\tau}$ to give to the subtask (it gives one partition of the resource type that would lead to the largest increase in execution rate).

If the function $\resourceAlloc$ cannot find such a task, then no more resources can be given 
and \algo skips the rest of this step (Line~\ref{line:18}) and moves to Step 3. Otherwise, it will 
add the extra budget $\delta_{\tau}$ to $\tau$'s current budget $\beta_{\tau}$ (Line~\ref{line:19}) and 
recalculate the completion time, and shorten the deadline of $\tau$ by an amount equal to the reduction in execution time under the newly increased budget (Line~\ref{line:20}). 

The completion time, which is used to define future decision points, is computed by the $\finishTime$ function, 
shown in Algorithm~\ref{alg:computed}. It works by considering all the phases $\theta_j$ that $\tau$ would execute under the current budget, starting from the current instruction count. It computes the total time that $\tau$ will spend in these phases, assuming that $\tau$ would be given the current budget $\beta_{\tau}$ up to the next decision point and  its base budget $\beta_{\tau}^{init}$ afterwards. The time spent at each phase 
 $\theta_j$ is calculated based on the execution rate $\theta_j^r$ and the number of instructions in $\theta_j$ that $\tau$ will execute. 

If $\tau$ (the task chosen to receive additional resources) is not in the scheduled set $S$, \algo will check whether it should swap the scheduled subtask $\tau^s$ 
with the latest deadline in $S$ with $\tau$; this happens if $\tau$ has a smaller deadline than $\tau^s$
and the swap does not lead to the scheduled subtasks having more total budget than $R_{\max}$ 
(Lines~\ref{line:21}--\ref{line:24}).  

Next, \algo checks whether $\tau$'s new completion time is no earlier than the next decision point $\tnext$ (Line~\ref{line:25}). If so, it simply recalculates the scheduled set $S$ and the total budget $R_s$ used by $S$, which is necessary because $\tau$'s deadline and resource allocation have changed (Line~\ref{line:26}). Otherwise, 
we have a newly created future decision point (at $c_{\tau}$) that is earlier than the immediate next decision point $\tnext$. In that case, it will invoke the function $\resetSegment$ to reset the 
resource-deadline allocation at the current decision point $t$ for all subtasks in $Q$  {\em except} for the subtask $\tau$ (Line~\ref{line:28}). It then updates $\tnext$ to be $c_{\tau}$ (Line~\ref{line:tnext-reset}).  This function resets the current budget $\beta_{\tau'}$ to be the base budget $\beta_{\tau'}^{init}$ 
and recomputes the deadline $d_{\tau'}$ under the new budget for every ready subtask $\tau' \neq \tau$. 
The intuition behind this reset is that, since $c_{\tau} < \tnext$, the current allocation for the subtasks, which is to be applied for the timing segment from $t$ to $t_{\mathsf{next}}$, may not be the most resource efficient if it is used instead for the shorter timing segment $[t, c_{\tau})$. This is because the improvement in the execution rates (and thus WCETs), which we aim to maximize in our resource allocation, varies depending on the timing segment we are computing the schedule for. It is influenced by where (which phase) in the program each subtask is currently in, how many instructions until the next phase, and how much improvement in execution rate the subtask would get if it is given extra resources at the current phase. For this reason, we redo the allocation for the subtasks to avoid being stuck in an inefficient allocation. Note that we only reset the allocation for other subtasks in $Q$ but not $\tau$, since we successfully shrink $\tau$'s completion time by giving it $\delta_{\tau}$ extra budget. Because of the changes in the subtasks' deadlines under the new resource budget, the function  $\resetSegment$ also computes and returns a new scheduled set $S$ and its total budget $R_s$. 

After obtaining the new $(S, R_s)$, \algo continues to the next iteration of resource allocation (repeat from Line~\ref{line:17}).
It will keep allocating resources until no more resources can be given, at which point it will move to Step 3.

{\bf Step 3 (Lines~\ref{line:29}--\ref{line:tnext-update}):} 
Once the algorithm cannot give out any more resources to subtasks, the scheduled set $S$ is fixed for the current decision time point $t$.
\algo will reset the current resource budget for all ready subtasks that are not in the scheduled set to be their base budget and reset to their initial deadlines (Line~\ref{line:30}). It then computes their new completion times, given that they were not scheduled in the current segment (Line~\ref{line:31}). \algo then traverses through all subtasks in the scheduled set $S$ (Line~\ref{line:32}). For each scheduled subtask $\tau$, it uses the function
\computeRetiredInstruction to compute the number of instructions $\Retired_{\tau}$ that $\tau$ would have completed by $\tnext$ (i.e., after executing in the segment $[t, \tnext)$) under its current allocated budget (Line~\ref{line:33}).
If $\Retired_{\tau}$ is greater than or equal to $\tau$'s total number of instructions, then $\tau$ would have finished its execution by $t_{\mathsf{next}}$; hence, we set $\tau$'s completion time to be $\tnext$ and $\done_{\tau}$ to true to indicate that $\tau$ has completed at the next decision point (Line~\ref{line:34}).
\algo will then shift the release times and completion times of all successor subtasks of $\tau$ based on the completion of $\tau$ (Line~\ref{line:35}), and add any ready successor subtasks to 
$Q$ while removing $\tau$.

Finally, $\algo$ saves the current allocation for the current decision point $t$ to the output schedule (Line~\ref{line:36}). 
If the ready set $Q$ is empty and there is no more future anchor point, then the static schedule is completed (Line~\ref{line:38}) and the algorithm moves to check for schedulability (Line~\ref{line:41}) before returning. 

Otherwise, \algo prepares data structures for the allocation in the next segment, starting from the new current decision point $t$ (Line~\ref{line:39}). It marks all source subtasks whose release times are equal to $t$ as ready and adds them to the ready set $Q$, then updates the set of future anchor points $B$ (Lines~\ref{line:release-anchor}--\ref{line:B-update}). 
Finally, it updates the immediate next decision point after $t$ to be the minimum of any completion time of the subtasks in the ready queue or the earliest future anchor point (Line~\ref{line:tnext-update}). The algorithm then continues onto the next iteration to compute the schedule for the next timing segment $[t, \tnext)$ (restarting from Step 1 onwards). \\[-2ex]

\noindent{\bf Schedulability test and output (Lines~\ref{line:41}--\ref{line:43}):} 
Once the schedule is completed ($B$ is empty), the algorithm will check for schedulability. The system is schedulable if every subtask in $J$ completes execution before the deadline of its corresponding task instance  (Lines~\ref{line:41}--\ref{line:43}). The algorithm returns the schedulability status and outputs the schedule $\Sched$.

\vspace{-1ex}
\section{Evaluation}\label{sec:eval}
\noindent
To evaluate the effectiveness of our generative profiling and resource-deadline co-allocation algorithms, we conducted a series of experiments using synthetic real-time DAG tasks and resource-intensive benchmarks as workloads. Our goals were to evaluate 
(1) the accuracy of using multi-marginal Schrödinger bridges to compute generative profiles, 
(2) the schedulability performance of \algo compared to the state of the art, and 
(3) the running time of \algo.

\begin{figure*}[t!]
    \centering
    \begin{subfigure}[b]{0.490\textwidth}
        \centering
        \includegraphics[width=\textwidth]{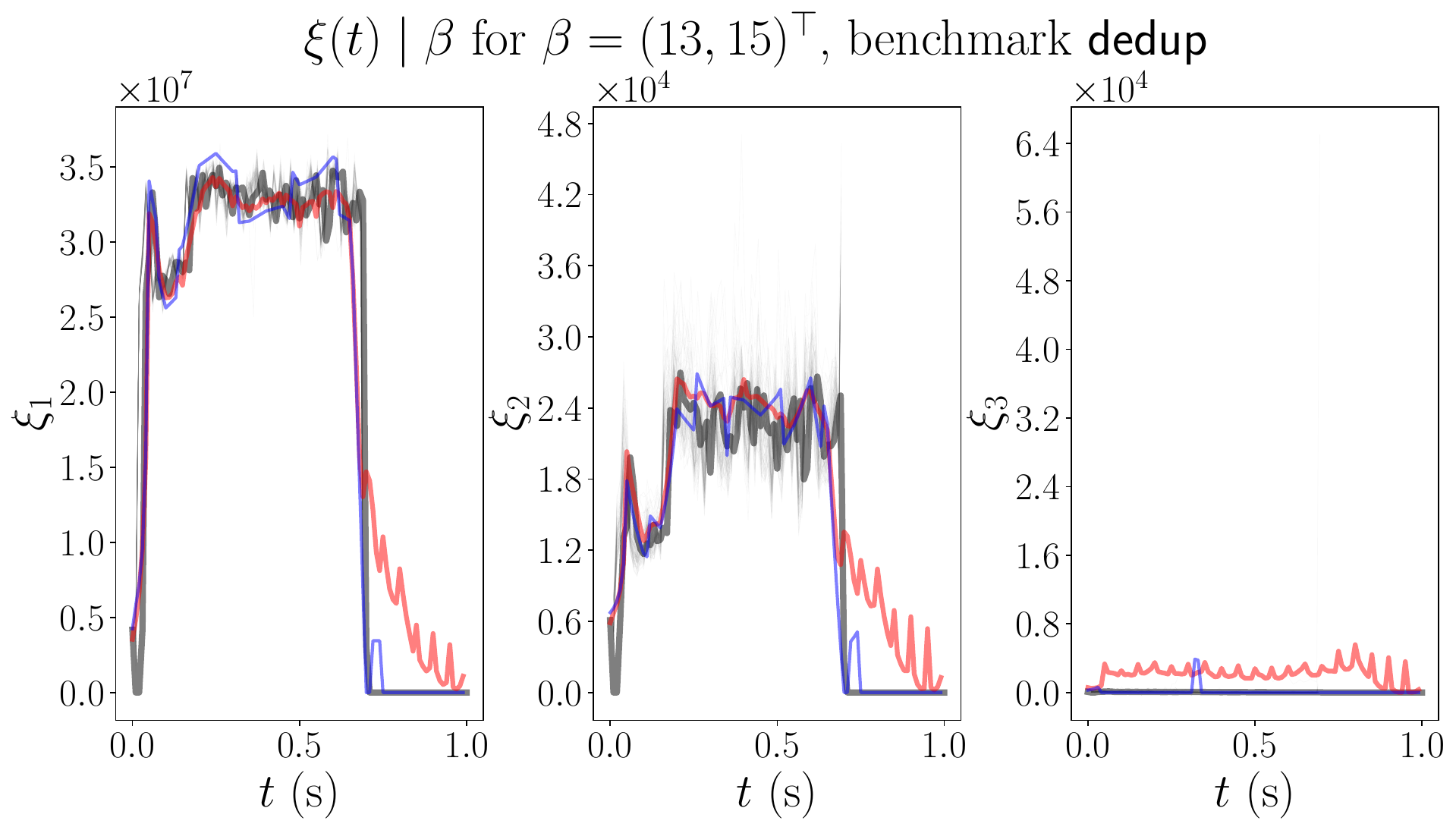}
    \end{subfigure}
    \hfill
    \begin{subfigure}[b]{0.490\textwidth}  
        \centering 
        \includegraphics[width=\textwidth]{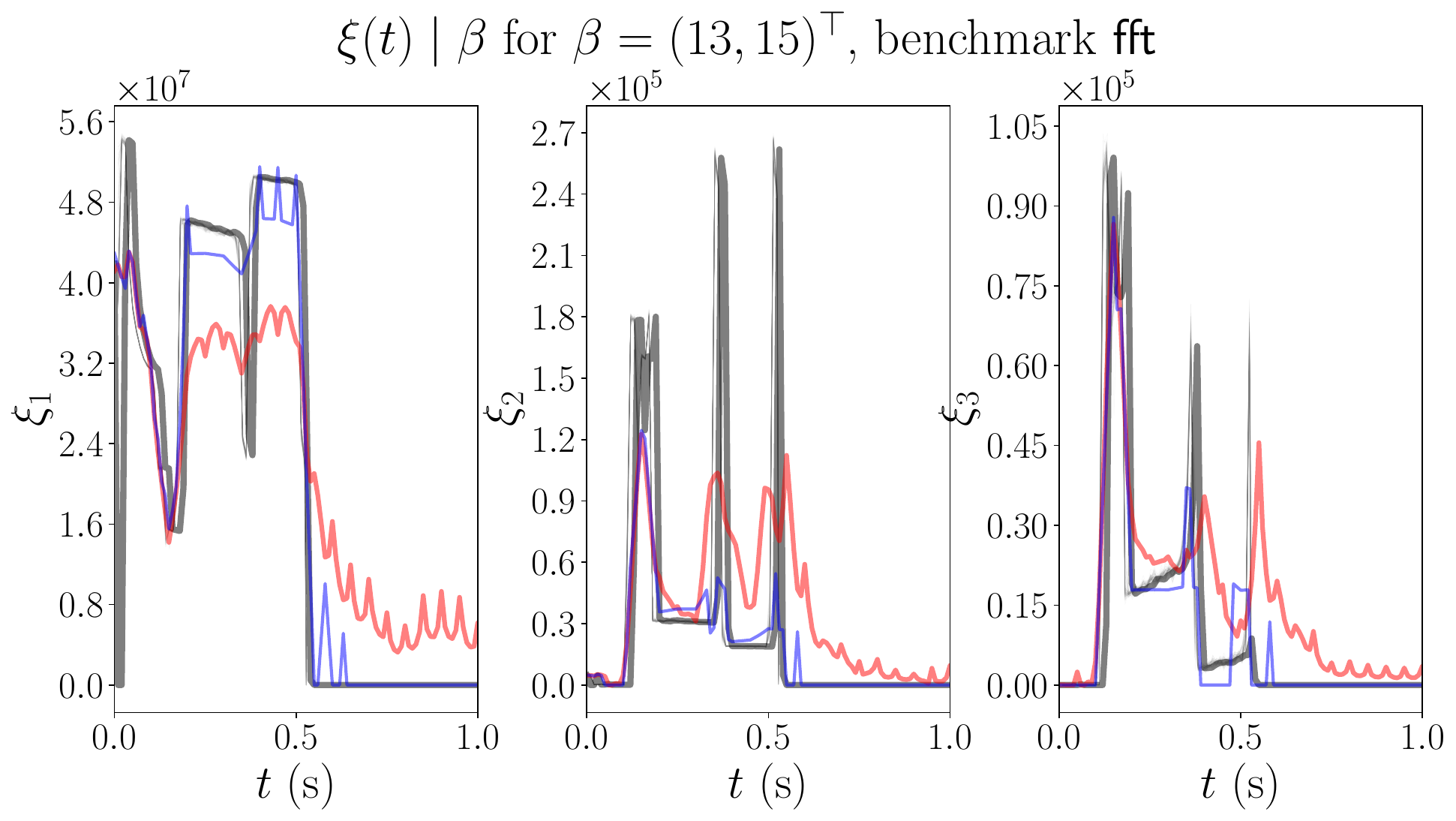}
    \end{subfigure}
    \vskip\baselineskip
    \vspace*{-0.1in}
    \begin{subfigure}[b]{0.490\textwidth}   
        \centering 
        \includegraphics[width=\textwidth]{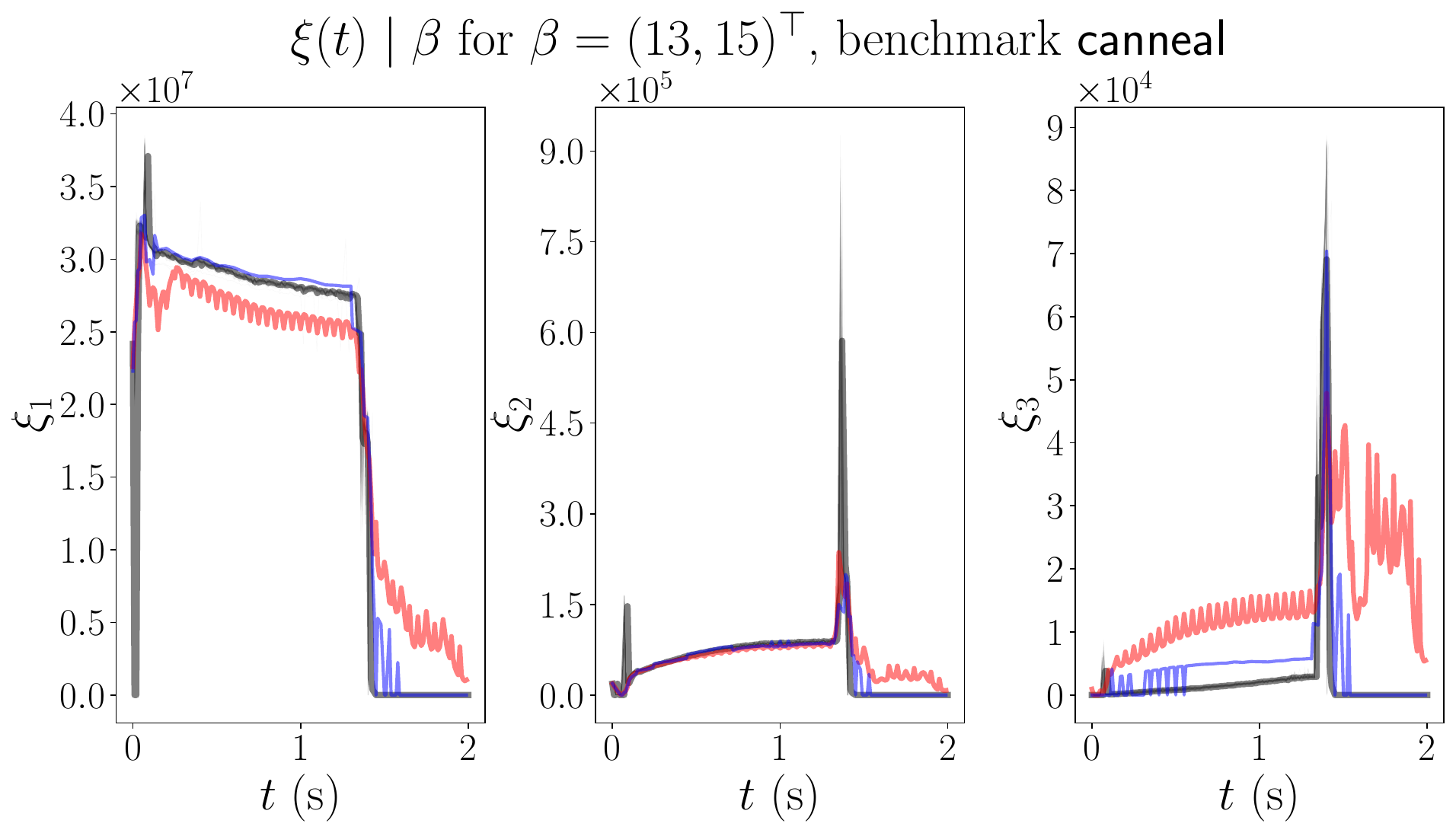}
    \end{subfigure}
    \hfill
    \begin{subfigure}[b]{0.490\textwidth}   
        \centering 
        \includegraphics[width=\textwidth]{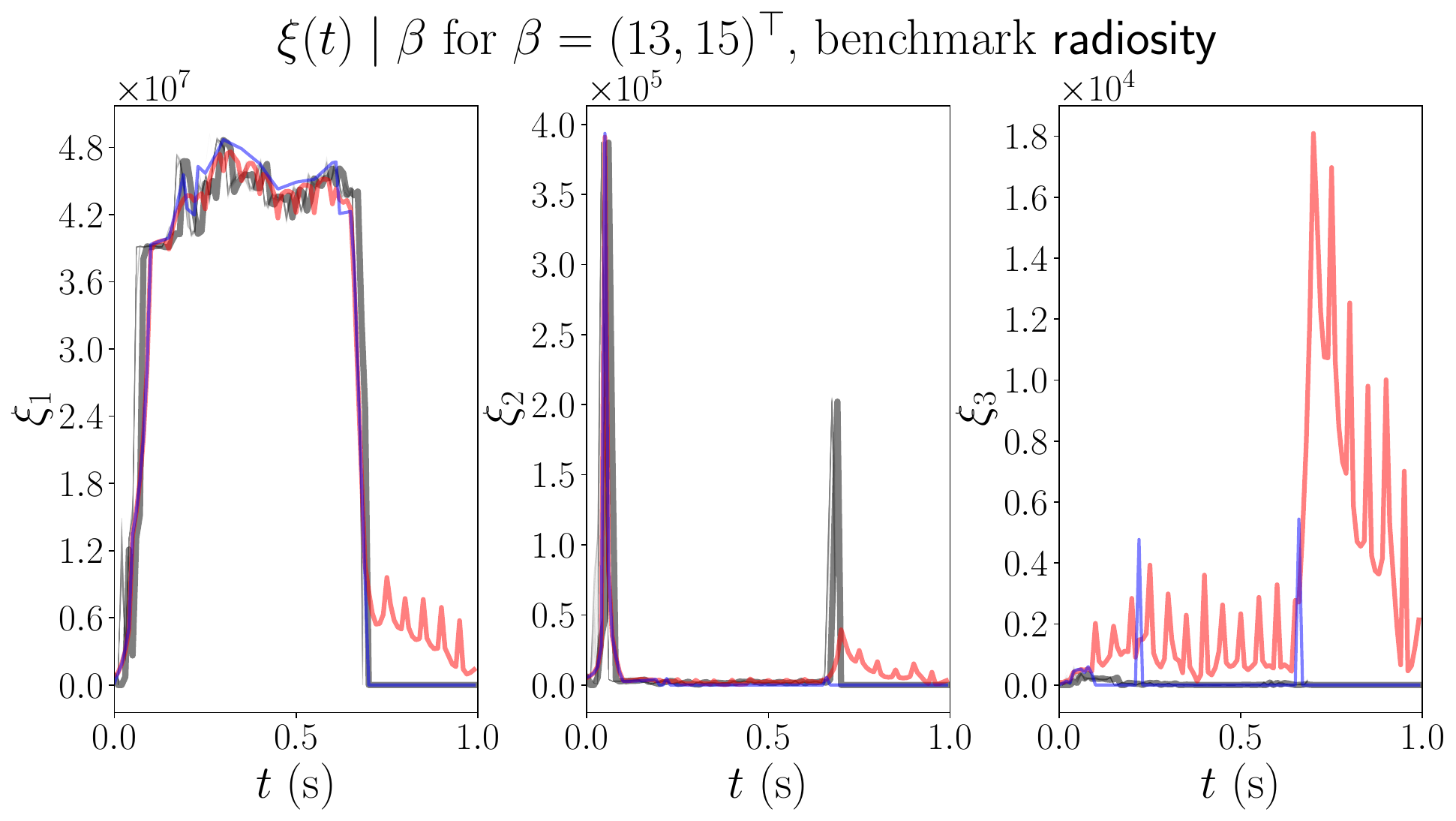}
    \end{subfigure}
    \vspace*{-0.05in}
    \caption{{\small{Maximum-likelihood synthetic profile ({\color{blue}blue}), mean synthetic profile ({\color{red}red}), mean empirical profile (black), and all empirical profiles ({\color{gray}grey}) for the benchmarks $\mathsf{dedup,\: fft,\: canneal,\:}$ and $\mathsf{radiosity}$ when $\beta=(13,\:15)^\top$.}}}
    \label{FigSynthProfiles_OneContext}
    \vspace{-2ex}
\end{figure*}

\noindent{\bf Workload measurement.}
We evaluated our generative model's accuracy and \algo's schedulability on a collection of workloads with varying resource needs. Our workloads include four 
popular benchmarks from the PARSEC and SPLASH-2x benchmark suites~\cite{PARSEC08}~\cite{splash2x}: $\mathsf{dedup,\: fft,\: canneal,\:}$ and $\mathsf{radiosity}$.
To gather the empirical data used as input to our model, we ran each workload on an Intel Xeon E5-2683 v4 processor with a 40 MB L3 cache.
We used Intel's CAT~\cite{intelCAT} to divide the cache into $N_{ca} = 20$ partitions. Using the technique from~\cite{Yun16-memguard-journal}, we measured a guaranteed memory bandwidth of 1.4 GB/s, which we divided into $N_{bw} = 20$ partitions of 70 MB/s each using MemGuard~\cite{Yun16-memguard-journal}.  Using these mechanisms to set a resource allocation, we ran each workload to completion under each possible resource budget, and recorded data for 100 runs. In each run, we recorded the number of instructions retired, cache requests, and cache misses ($\xi_1, \xi_2, \xi_3$, respectively in our model) once every 10ms.

\subsection{Accuracy of generative stochastic model}
\noindent Given this empirical data, we used the approach outlined in Section~\ref{sec:prediction} to generate `synthetic' resource profiles conditioned on $\beta$ (i.e. $\xi(t)\mid\beta$). For the experiment, we restricted the number of resource allocations $\beta$ and the number of profiles $n_d$ available to our MSBP solver. Specifically, we had 
$$ \mathcal{B} = \llbracket N_{\mathsf{ca}}\rrbracket\times \llbracket N_{\mathsf{bw}}\rrbracket \quad,\quad \mathcal{B}' = \{1,5,10,15,20\}^2 $$
and $n_d=10$, meaning we took 10 random profiles from the 100 available. From these limited profiles, we constructed the marginal distributions \eqref{defmusigmaempirical} for times $t_\sigma=0.05\cdot(\sigma-1)$, with $t_{n_s}$ chosen depending on the approximate runtime of the benchmark. As $n_b=|\mathcal{B}'|=25$, each $\mu_\sigma$ then had $n_dn_b=250$ scattered data points.

\begin{remark}
    The restriction of the number of profiles $(n_d)$ and the number of allocations $(n_b)$ provided as input to the MSBP solver has a practical motivation -- if either the number of workloads to profile or the number of allocations $|\mathcal{B}|$ is large, it is intensive both in time and computation to empirically generate a large number of profiles for all workloads and for each resource allocation. Our method allows for a significant reduction in profiling time as it requires only a small subset of $\mathcal{B}$ and a small $n_d$ to generate synthetic profiles for all $\beta\in\mathcal{B}$. In this particular case, a total number of $n_dn_b=250$ profiles were needed to be generated for each benchmark, as opposed to $|\mathcal{B}|\cdot100=40,\!000$. Moreover, the entire process of MSBP solution and synthetic profile generation for all $\beta\in\mathcal{B}$ took $\approx$15 minutes per benchmark on consumer-grade computing hardware.
\end{remark}

Solving the MSBP over the marginals $\{\mu_{\sigma}\}_{\sigma\in\llbracket n_s\rrbracket}$, we generated the most likely conditional joint distributions $\mu_t/\int_\mathcal{X}\mu_td\xi$ per \eqref{ComputeConditional} at times from 0 to $t_{n_s}$ every 10ms, and for each $\beta\in\mathcal{B}$. As each of these joints are \emph{weighted} scattered distributions, we obtained from their aggregate a synthetic resource profile of the benchmark conditioned on $\beta$ by taking $\xi(t)\mid\beta$ to be the data point with the highest probability value in the respective conditional joint distribution, for each $t\in\{0,0.01,0.02,\dots,t_{n_s}\}$. We call this the \emph{maximum-likelihood} synthetic profile.

To visualize the `accuracy' of these synthetic profiles, for each $\beta\in\mathcal{B}$ we averaged out the 100 empirical profiles to obtain a mean empirical resource usage profile, which could then be compared to our synthetic profile for the same $\beta$. Fig. \ref{FigSynthProfiles_OneContext} shows the overlaid profiles for $\beta=\left(13,15\right)^\top$ for the various benchmarks.

\begin{remark}
    It may seem natural to choose the \emph{mean} value of each conditional joint distribution to construct our synthetic profiles. Thus, Fig. \ref{FigSynthProfiles_OneContext} shows synthetic profiles constructed in this way as well. As can be seen most clearly in the `tail-end' behavior of such profiles, however, this choice will lead to inaccurate results when $\mu_t$ is nongaussian (particularly, multimodal). As the MSB allows us to obtain the \emph{maximum-likelihood} distributions $\mu_t$ in a nonparametric manner, we can obtain accurate results without assumptions on the nature of these distributions.
    \label{rmk:maxlikelihood_vs_mean}
\end{remark}

Fig. \ref{FigSynthProfiles_OneContext} shows that for all benchmarks, our synthetic profiles capture the time-varying resource usage patterns of the software, even for contexts without the training set $\mathcal{B}'$. It is difficult to meaningfully gauge the absolute accuracy of our synthetic profiles in a quantitative manner, since the evaluation of any metric on the space of curves in $\mathbb{R}^2$ will be dependent on the magnitude of the curves themselves. We therefore eschew such comparisons and instead use these synthetic profiles as input to \sys in subsection~\ref{sched-eval}, where we compare scheduling performance to the case where we use empirical profiles.

\begin{figure*}[t!]
    \centering
    \begin{subfigure}[b]{0.32\textwidth}
        \centering
        \includegraphics[width=\textwidth]{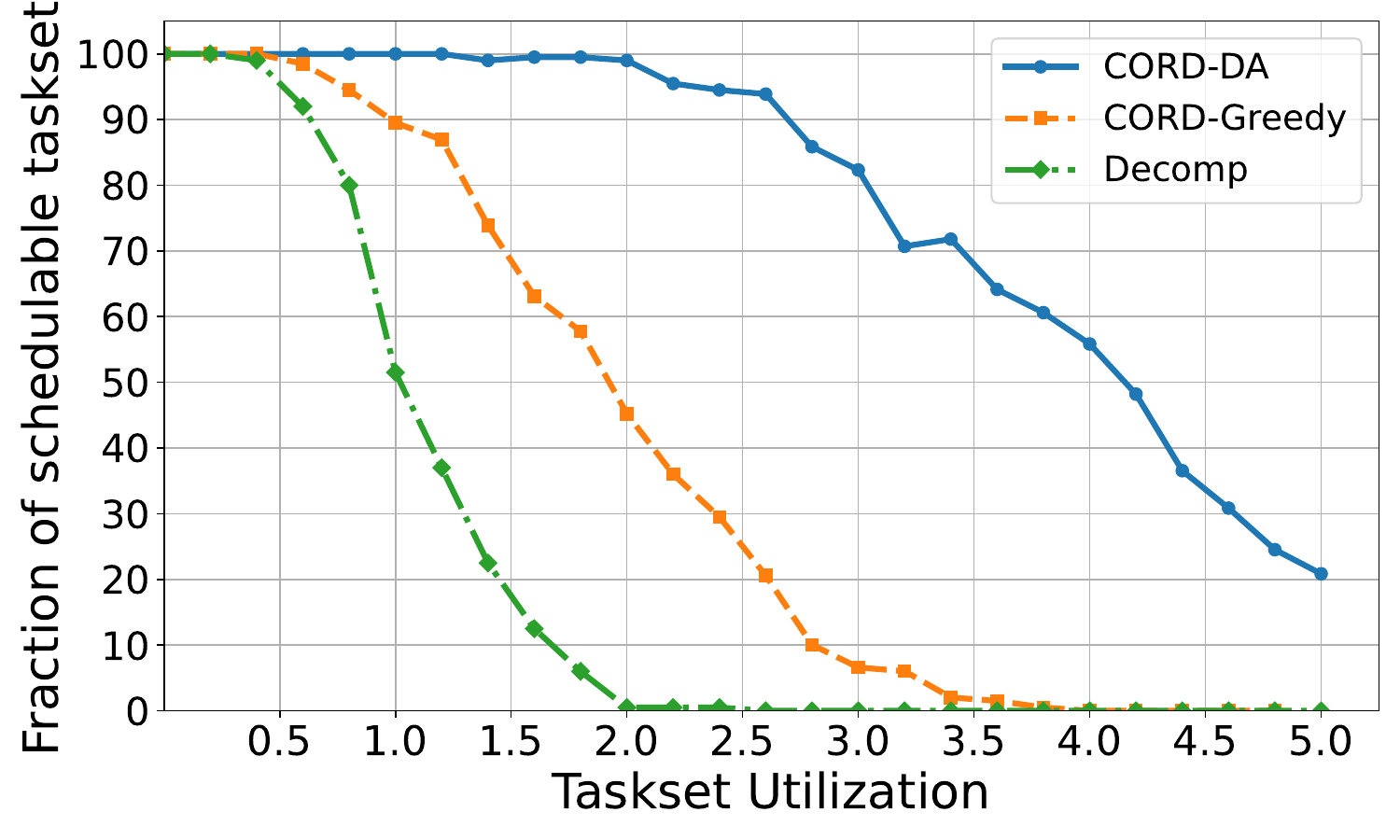}
        \caption{$p = .25$}  
        \label{fig:subfig1}  
    \end{subfigure}
    \hfill
    \begin{subfigure}[b]{0.32\textwidth}  
        \centering 
        \includegraphics[width=\textwidth]{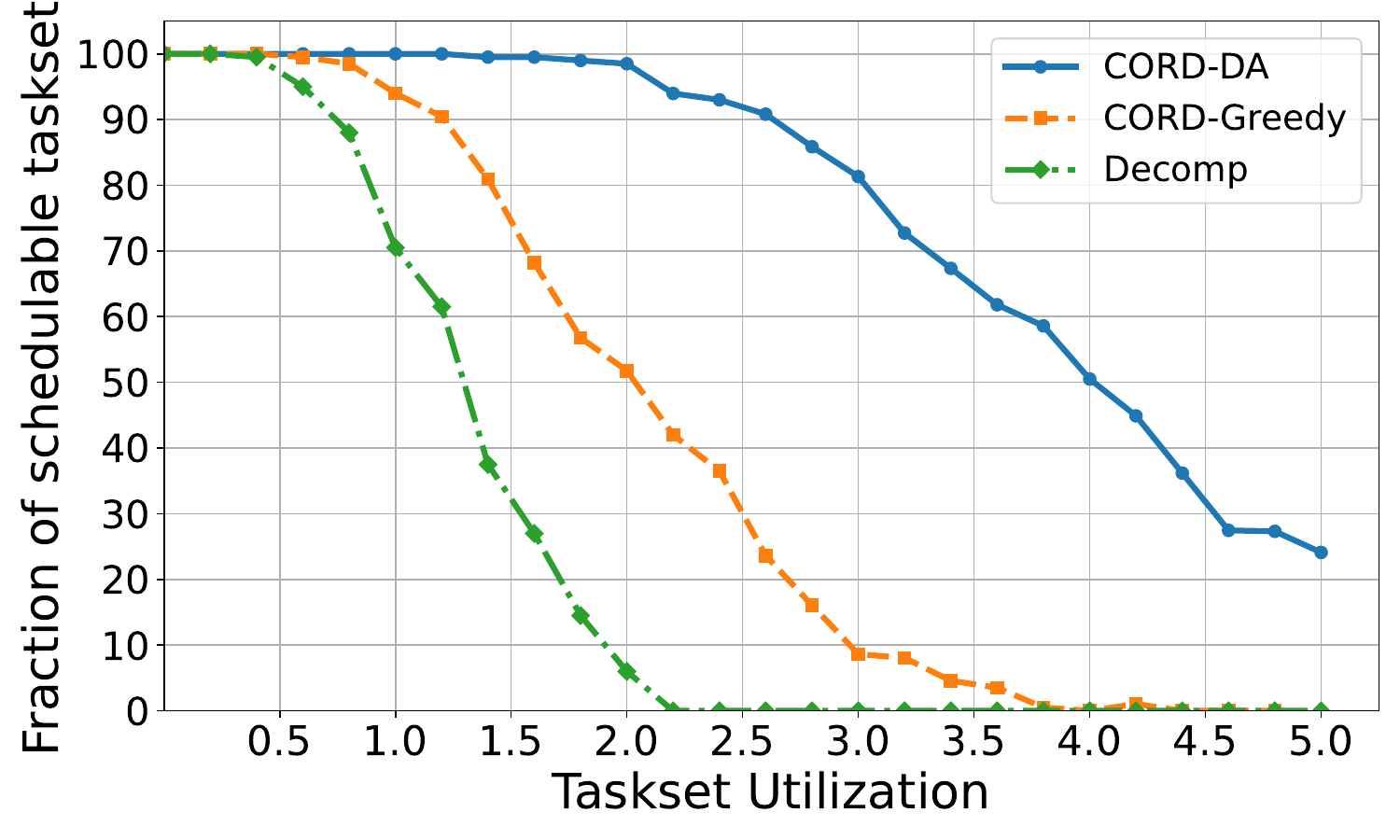}
        \caption{$p = .50$}  
        \label{fig:subfig2}  
    \end{subfigure}
    \hfill
    \begin{subfigure}[b]{0.32\textwidth}   
        \centering 
        \includegraphics[width=\textwidth]{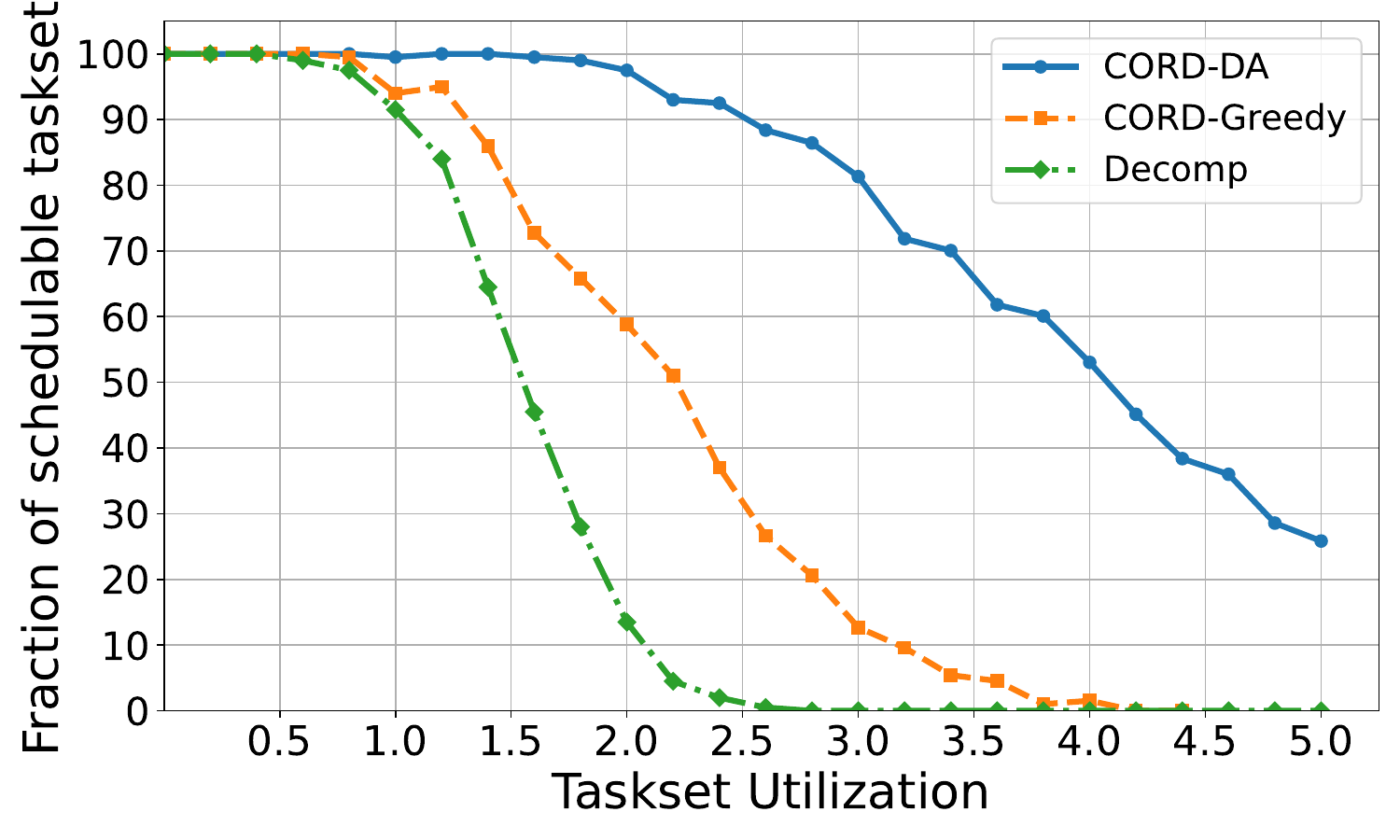}      
        \caption{$p = .75$}  
        \label{fig:subfig3}  
    \end{subfigure}
    \vspace*{-0.05in}
    \caption{Fraction of schedulable tasksets on $m = 4$ cores for different $p$ values}  
    \label{Fig:schedulability_4core}  
    \vspace*{-0.1in}
\end{figure*}

\subsection{Schedulability performance of \algo}\label{sched-eval}
\noindent
\noindent{\bf{Taskset generation.}}
To evaluate \algo's effectiveness at scheduling DAG-based tasks, we generated DAG tasksets using a similar approach to~\cite{singlefp2}. 
We leveraged their DAG creation tool~\cite{xiaotian_dai_2022_6334205} to generate 200 unique tasksets, with taskset utilization ranging between 0 and 5.0, at steps of 0.2. This produced a total of 5,000 tasksets per experiment. 

Each taskset consists of 5  DAG tasks with utilizations $U_{\tau}$ falling uniformly in the range [0, 1], which are calculated by the classic UUniFast-Discard algorithm~\cite{Emberson2010a}.
The end result is a DAG task whose subtasks' total utilization is the taskset's target utilization. Each DAG has a minimum depth of 3 and a maximum depth of 8 layers, respectively. Each vertex (subtask) in the DAG has probability $p$ of connecting to another vertex in a layer below it.
We used the tool to limit the number of vertices that can appear in any single layer to 4.

We next assigned workloads to each vertex (subtask), picked uniformly at random from our set of benchmarks. The execution time $e_{\tau | \beta^r}$   of each subtask $\tau$ is the WCET under uniform resource allocation  
$\beta^r =  (N_{\mathsf{ca}}/m,N_{\mathsf{bw}}/m)$ of its workload. We refer to this execution time as the \textit{reference} execution time.
We then calculated a period for the DAG task such that the utilization assigned from the DAG creation tool is maintained: $P = \frac{\sum e_{\tau | \beta^r}}{U_{\tau}}$. 
Lastly, we rounded the period to the nearest harmonic period defined by the closest power of 2. While this makes the resulting utilization of the DAG task graph deviate from the original target utilization, it enables us to calculate manageable hyper-periods for our tasksets. \\[-2ex]

\noindent{\bf Implemented algorithms.}
We implemented both the greedy and deadline-aware approaches for computing the initial base resource budgets used in \algo. We refer to the two variants of \algo as \algo-Greedy and \algo-DA, respectively.
For comparison, we also implemented a recent deadline decomposition scheduling algorithm from~\cite{stretch}, referred to as \baseline.
Similar to \algo, \baseline computes deadlines and release times for subtasks and additionally provides a schedulability analysis to determine if the computed values are schedulable. Unlike \algo, however, \baseline  does not consider resource allocation. Hence, for our experiments, we assumed that it is used with a static even allocation of resources across all cores. We used the analysis for GEDF in~\cite{stretch} to test the schedulability under \baseline. 
All implementations of the three algorithms, \algo-Greedy, \algo-DA and \baseline, within 1490 lines of Python code and 617 lines of C code (approximately 2,100 LOCs in total).\\[-2ex]


\begin{figure*}[t!]
    \centering
    \begin{minipage}[b]{0.32\textwidth}
        \centering
        \includegraphics[width=\textwidth]{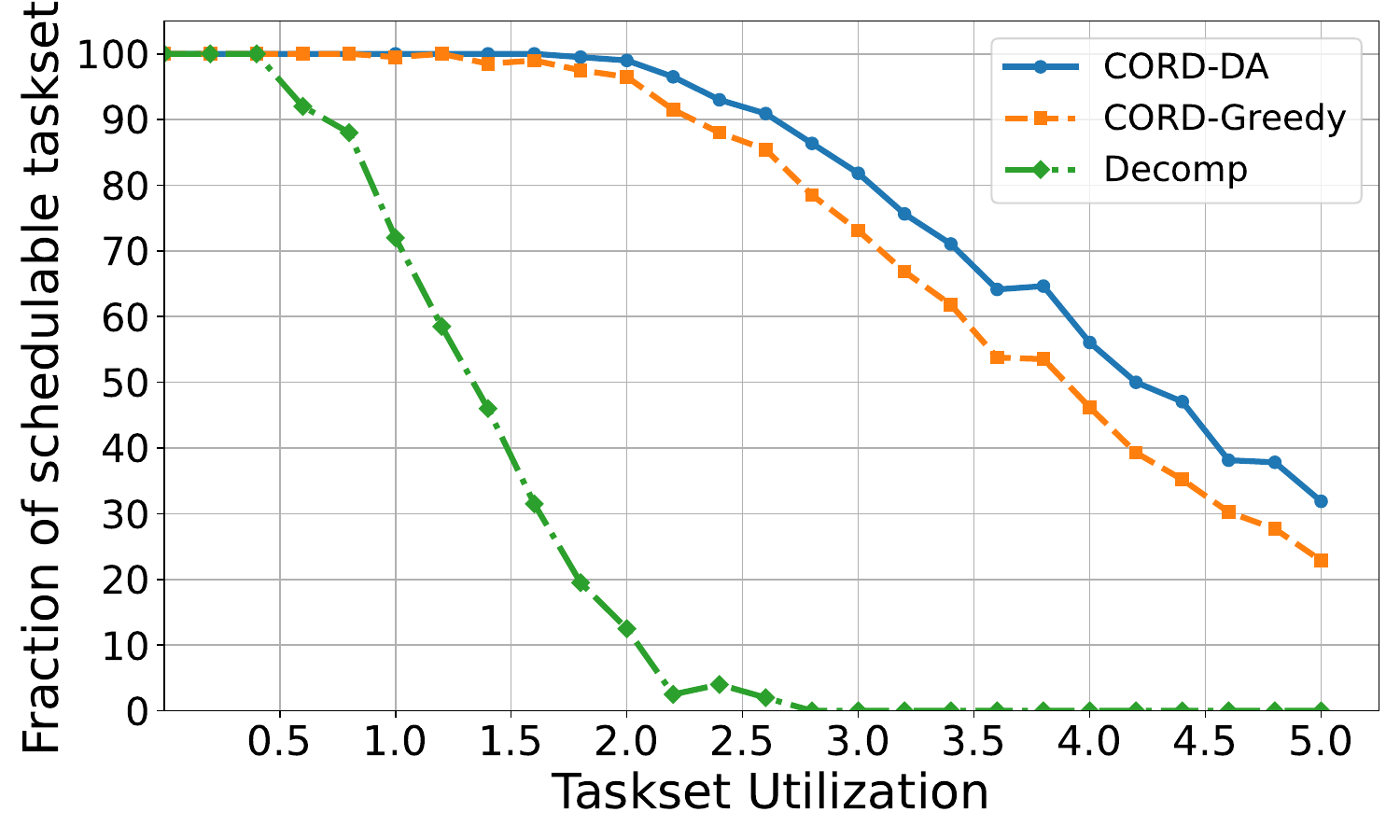}
        \caption{Schedulability ($m = 10$, $p = 0.5$)}  
        \label{Fig:schedulability_10core}  
    \end{minipage}
    \hfill
    \begin{minipage}[b]{0.32\textwidth}  
        \centering 
        \includegraphics[width=\textwidth]{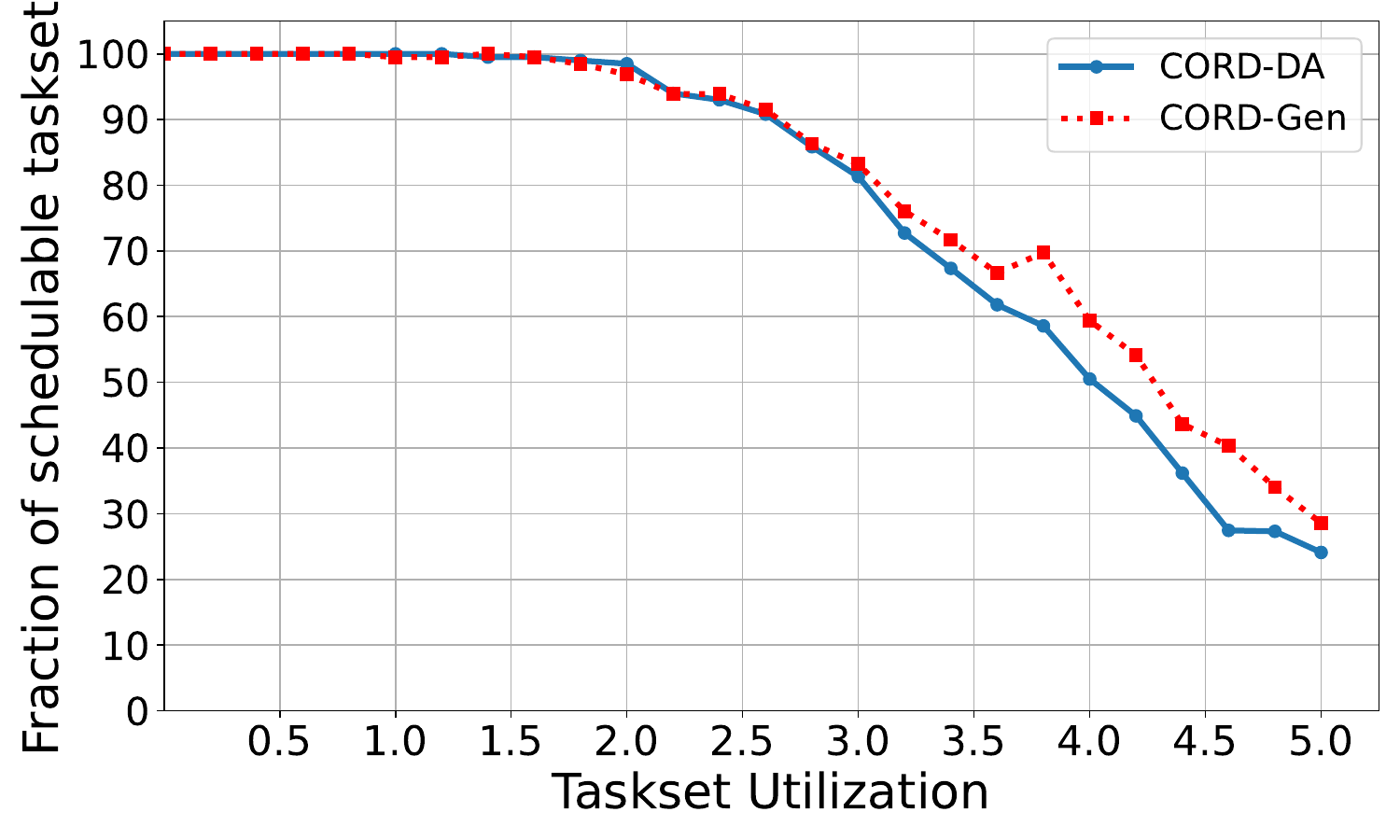}
        \caption{\algo with generative profiles.}  
        \label{Fig:schedulability_plot_synth}  
    \end{minipage}
    \hfill
    \begin{minipage}[b]{0.32\textwidth}\centering
        \raisebox{2.2cm}{
            \scalebox{0.8}{
                \begin{tabular}{|c|c|c|}
                \hline
                              & \algo-DA       & \algo-Greedy   \\ \hline
                Min           & 0.196                   & 1.950                       \\ \hline
                Max           & 383.4                   & 587.6                       \\ \hline
                \textbf{Avg.} & \textbf{11.33}          & \textbf{40.02}              \\ \hline
                99th          & 160.7                   & 279.9                       \\ \hline
                \end{tabular}
            }
        }\vspace{-4ex}
        \captionof{table}{Running time (s).}\label{tab:runtime}
    \end{minipage}
\label{tab:runtime}
    \vspace{-3ex}
\end{figure*}

\noindent{\bf Schedulability results using empirical profiles.}
We first demonstrate \algo's ability to schedule tasksets for a variety of DAGs without our generative profiles. By using the complete set of empirical profiles,
we can better assess the standalone performance of our co-allocation algorithm. 

We considered a range of diverse DAG structures by varying the vertex's  probability $p$ value in $\{0.25, 0.5, 0.75\}$, representing a 25\%, 50\% and 75\% chance that vertices have an edge between them, respectively. 
Fig.~\ref{Fig:schedulability_4core} shows the schedulability results on a 4-core platform. 

As expected, as the taskset utilization increases, the fraction of schedulable tasksets also decreases for all algorithms. As we increase $p$, the tasksets become slightly easier to schedule due to the increasingly sequential DAG-structures. However, across all three values of $p$, both variants of our algorithm observe significant improvements in schedulability compared to \baseline. Both \algo-DA and \algo-Greedy are able to schedule much heavier tasksets, due to their more effective use of the shared  resources and their resource-deadline co-design. For instance, with $p = .25$,  \baseline does not identify any tasksets as schedulable (on four cores) even at utilization 2.0. In contrast, \algo-DA can schedule all of the tasksets at 2.0 utilization, and more than half of the tasksets  at 4.0 taskset utilization. We also observe that \algo-DA consistently outperforms \algo-Greedy, which is expected because the latter does not consider tasks' deadlines in its initial allocation. 

Next, we evaluated how well \algo scales across platforms with more cores.
Figure~\ref{Fig:schedulability_10core} shows the schedulability results for 
tasksets with $p=0.5$ on a platform with $m = 10$ cores.
As discussed above, we provided \baseline with an even split of resources (i.e., 2 cache partitions and 2 memory bandwidth partitions per core.) 
The results are consistent with the observations for four cores: both \algo-Greedy and \algo-DA continue to substantially outperform \baseline. 
Note that the performance gap between \algo-DA and \algo-Greedy tightens when the tasksets are given additional cores.
The extra cores allow \algo-Greedy to maximize resource efficiency, without being negatively impacted by its greedy deadline choices.
This is expected, as having more cores makes the algorithm less reliant on intelligent deadline allocations.
However, \algo-DA performs about the same, given that 1) the available resource budget did not change and 2) regardless of the number of cores, \algo-DA chooses subtask deadlines that provide as much slack as possible, such that total deadlines are met. \\[-2ex]



\noindent{\bf Running time.}
Table I shows the running time statistics of \algo-DA and \algo-Greedy for an experiment with 4 cores and $p = 0.5$, computed over all utilizations.
Interestingly, we observe that \algo-DA had a lower running time compared to \algo-Greedy across all cases. This seems surprising, as our deadline-aware version  has  additional preprocessing steps to calculate deadlines and release times.
However, upon close inspection, we observed that \algo-Greedy spent far more time swapping subtasks in and out of its set of scheduled subtasks (Lines 24--29 in Algorithm~\ref{alg:cord}) and resetting its allocation, which led to a higher runtime. Overall, the running time of both variants of our algorithm is reasonable. \\[-2ex]


\noindent{\bf Schedulability using generative profiles.}
Lastly, we evaluated how well \algo performs when it uses the generative profiles. Recall that in our experiments, the `synthetic' generative profiles were generated using our generative stochastic model that trained on only a small subset (10\%) of the empirical profiles. We then constructed the multi-phase model for the subtasks from only these generated profiles. The constructed phases were then used by \algo-DA. We call this generative-based version  \algo-Gen. By comparing \algo-Gen with the fully-empirical \algo-DA (which uses the complete set of empirical profiles, taking days to collect), we can assess whether our generative profiles are beneficial in improving schedulability and how well it can predict resource characteristics even with very limited empirical data. 

We evaluated \algo-Gen using the same tasksets we used to evaluate \algo-DA in Figure~\ref{fig:subfig2}. Its schedulability result, together with that of \algo-DA, are shown in Figure~\ref{Fig:schedulability_plot_synth}. 
Note that the schedulability performance is nearly identical between \algo-Gen and \algo-DA from early to mid taskset utilizations, despite the fact that only 10\% of the profiled data is used to train our model. We account for the increase in schedulability at higher utilizations due to the fact that the model's maximum likelihood method of generating profiles filters out outlier profile results that cause the rate to drop (due to system related overheads such as kernel interrupts). This leads to slightly faster execution rates from the synthetic data, resulting in improved schedulability at high utilizations.


\section{Conclusion}
\noindent 
We have presented a generative stochastic model for predicting software's 
resource profiles on multicore platforms and a resource-deadline co-allocation 
algorithm for DAG-based tasks that leverages the resource profiles
to maximize resource efficiency and schedulability.
Our generative model allows for an order of magnitude  
reduction in profiling time as it requires only a small subset of empirical data,
and our evaluation results show that it can accurately capture the 
time-varying resource usage patterns of the 
software, even for contexts without the empirical training set.
Using the profiles, we can construct a richer timing
model for the workload that exposes the time-varying impact of 
resource allocation on the workload's execution time.
By leveraging this information in making scheduling decisions,
and by jointly considering resource allocation with execution time and deadline, 
our co-allocation algorithm achieves substantially better 
schedulability and supports much heavier loads than 
the state of the art.


\newpage
\balance
{
\bibliographystyle{abbrv}
\bibliography{paper,abby}

}

\newpage

\onecolumn

\begin{center} \LARGE{\bf{Supplementary Materials}} \end{center}

\begin{multicols*}{2}
\section*{Multimarginal Schrödinger Bridge}
This section outlines the Multimarginal Schr\"odinger Bridge and its solution. We follow the exposition of \cite{bondar-2024-psmsbp}, where additional details may be found.\newline

Let $\{\mu_\sigma\}_{\sigma\in\llbracket n_s\rrbracket}$ be a set of `snapshots' -- probability distributions each consisting of $n_d$ scattered points $\{\eta^i(t_\sigma)\}_{i\in\llbracket n_d\rrbracket}$. In other words $\mu_\sigma$, viewed as a probability (column) vector, belongs to the standard simplex $\{\mu\in\mathbb{R}^{n_d}_{\geq 0}\mid \bm{1}^{\top}\mu=1\}$ where $\bm{1}$ denotes the all-ones vector of size $n_{d}\times 1$.

We use $\otimes$ as the tensor product symbol. We define $\bm{C}\in(\mathbb{R}^{n_d})_{\geq 0}^{\otimes n_s}$ as the tensor whose $i_1,\dots,i_{n_s}$th entry is the cost of transporting unit probability mass between $\eta^{i_1}(t_1),\dots,\eta^{i_{n_s}}(t_{n_s})$, i.e., the entries of the tensor $\bm{C}$ are
$$ [\bm{C}_{i_1,\dots,i_{n_s}}] = c(\eta^{i_1}(t_1),\dots,\eta^{i_{n_s}}(t_{n_s})) $$
where $c:(\mathbb{R}^{n_d})^{n_s}\to\mathbb{R}_{\geq 0}$ is some cost function. 

Let $\bm{M}\in(\mathbb{R}^{n_d})_{\geq 0}^{\otimes n_s}$ be a probability mass transport plan between the snapshots, where $[\bm{M}_{i_1,\dots,i_{n_s}}]$ is the probability mass transported between $\eta^{i_1}(t_1),\dots,\eta^{i_{n_s}}(t_{n_s})$. Any valid $\bm{M}$ must satisfy the constraint that the sum of all probability mass transported to the $j$th scattered point in the $\sigma$th snapshot must equal to $[(\mu_\sigma)_j]$. Equivalently, if we define the projection of $\bm{M}$ onto $\mu_\sigma$ (componentwise) as
\begin{equation}
    \Big[{\rm{proj}}_\sigma(\bm{M})_j\Big] := \sum_{i_1,\dots,i_{\sigma-1},i_{\sigma+1},\dots,i_{n_s}}\bm{M}_{i_1,\dots,i_{\sigma-1},j,i_{\sigma+1},\dots,i_{n_s}},\notag
\end{equation}
then this constraint can be expressed as
\begin{equation}
{\rm{proj}}_\sigma(\bm{M})=\mu_\sigma \quad\quad\forall\sigma\in\llbracket n_s\rrbracket,
\label{MSBP_constraint}
\end{equation}
wherein the mapping $${\rm{proj}}_\sigma:(\mathbb{R}^{n_d})_{\geq 0}^{\otimes n_s}\to\mathbb{R}_{\geq 0}^{n_d}.$$ 
Similarly, we may define the bimarginal projection of $\bm{M}$ onto its $(\sigma_1,\sigma_2)$th components as the mapping 
$${\rm{proj}}_{\sigma_1,\sigma_2}:(\mathbb{R}^{n_d})_{\geq 0}^{\otimes n_s}\to\mathbb{R}_{\geq 0}^{n_d\times n_d}$$ 
where
\begin{align}
    \Big[{\rm{proj}}_{\sigma_1,\sigma_2}(\bm{M})_{j,\ell}\Big] &:=\nonumber\\
&\!\!\!\!\!\!\!\!\!\!\!\!\!\!\sum_{\{i_\sigma\}_{\sigma\in\llbracket n_s\rrbracket\setminus\{\sigma_1,\sigma_2\}}}\!\!\!\!\!\!\!\!\!\!\!\!\!\!\bm{M}_{i_1,\dots,i_{\sigma_1-1},j,i_{\sigma_1+1},\dots,i_{\sigma_2-1},\ell,i_{\sigma_2+1},\dots,i_{n_s}}. \notag
\end{align}
If $\bm{M}$ satisfies \eqref{MSBP_constraint}, then the bimarginal projection is a scattered distribution supported over the finite set $\{\eta^i(t_{\sigma_1})\}_{i\in\llbracket n_d\rrbracket}\times\{\eta^i(t_{\sigma_2})\}_{i\in\llbracket n_d\rrbracket}$ with $\mu_{\sigma_1}$ and $\mu_{\sigma_2}$ as statistical marginals.

The Multimarginal Schr\"odinger Bridge Problem (MSBP) is to find the transport plan $\bm{M}$ with the minimal entropy-regularized probability mass transport cost, expressed as the optimization problem
\begin{subequations}
\begin{align}
&\underset{\bm{M}\in\left(\mathbb{R}^{n_d}\right)^{\otimes n_s}_{\geq 0}}{\min}~\langle\bm{C}+\varepsilon\log\bm{M},\bm{M}\rangle\label{DiscreteMSBPobj}\\
&\text{subject to}~~{\rm{proj}}_{\sigma}\left(\bm{M}\right) = \bm{\mu}_{\sigma}\quad\forall\sigma\in\llbracket n_s\rrbracket,\label{DiscereteMSBPconstr}
\end{align}
\label{DiscreteMSBP}
\end{subequations}
where $\langle\cdot,\cdot\rangle$ indicates the Hilbert-Schmidt inner product and $\varepsilon>0$ is the entropy regularization parameter. We refer to the minimizer of \eqref{DiscreteMSBP} as the Multimarginal Schr\"odinger Bridge (MSB) and denote it as $\bm{M}_{\rm{opt}}$.

What makes the MSBP \eqref{DiscreteMSBP} pertinent to learning from distributional data is a mathematical result from the theory of large deviations, specifically Sanov's theorem \cite{sanov1958probability}, \cite[Sec. II]{follmer1988random}. This theorem establishes that the minimizer of \eqref{DiscreteMSBP} is precisely the \emph{most likely} joint distribution subject to the observational constraints \eqref{DiscereteMSBPconstr} imposed at times where measurements are available. In other words, solving \eqref{DiscreteMSBP} is equivalent to solving a \emph{constrained maximum likelihood problem on the space of probability measure-valued curves traced over time}. Phrased differently, the solution of \eqref{DiscreteMSBP} is guaranteed to be statistically most parsimonious learning from distributional snapshot (i.e., scattered point cloud) data. 

The MSBP \eqref{DiscreteMSBP} is a \emph{strictly convex} program in $n_d^{n_s}$ decision variables. Thus, the existence-uniqueness of $\bm{M}_{\rm{opt}}$ is guaranteed. For computational tractability, it is standard to leverage Lagrange duality, whereby the MSB admits the structure 
\begin{align}
\bm{M}_{\rm{opt}}=\bm{K}\odot\bm{U},\quad \bm{K}=\exp(-\bm{C}/\varepsilon),\quad \bm{U}=\bigotimes_{\sigma\in\llbracket n_s\rrbracket}u_\sigma,
\label{Moptstructure}
\end{align}
wherein $\odot$ denotes elementwise (Hadamard) product, and the vectors 
$$u_\sigma:=\exp(\lambda_\sigma/\varepsilon)\in\mathbb{R}^{n_d}_{>0}$$ 
are the transformed Lagrange multipliers $\lambda_\sigma$ associated with the constraints \eqref{DiscereteMSBPconstr}. Note that the tensors $\bm{K},\bm{U}\in(\mathbb{R}^{n_d})_{\geq 0}^{\otimes n_s}$. 

By strong duality, we only need to solve for the exponential-transformed dual variables $\{u_\sigma\}_{\sigma\in\llbracket n_s\rrbracket}$, which are obtained by the Sinkhorn iterative scheme
\begin{align}
{u}_{\sigma} \leftarrow {u}_{\sigma} \odot {\mu}_{\sigma}\oslash{\rm{proj}}_{\sigma}\left(\bm{K}\odot\bm{U}\right) \quad\quad\forall\sigma\in\llbracket n_s\rrbracket .
\label{MultimarginalSink}    
\end{align}
In \eqref{MultimarginalSink}, the symbol $\oslash$ denotes elementwise (Hadamard) division. The Sinkhorn recursions \eqref{MultimarginalSink} are strictly contractive over the cone $\mathbb{R}^{n_d}_{>0}$, and therefore guaranteed to converge to a unique fixed point with linear rate of convergence.

Furthermore, in \eqref{MultimarginalSink}, efficient computation of the projection is possible by exploiting the structure of $\bm{C}$, which dictates the structure of $\bm{K}$, and hence the structure of $\bm{K}\odot\bm{U}$; see \cite[Proposition 1]{bondar-2024-psmsbp}. In the context of this work, the snapshots $\mu_\sigma$ are indexed by time ($t_\sigma$), which imposes a natural `path' structure to the cost tensor $\bm{C}$, i.e.,
$$ [\bm{C}_{i_1,\dots,i_{n_s}}] = \sum_{j=1}^{n_s-1}[C^j_{i_{j},i_{j+1}}],$$
where the $C^j\in\mathbb{R}^{n_d\times n_d}$ are matrices whose $(i,j)$th entry is the value $c(\eta^{i_j}(t_j),\eta^{i_{j+1}}(t_{j+1}))$ for some cost function $c$. In this work, we use the squared Euclidean cost $c$.

The Sinkhorn-recursion \eqref{MultimarginalSink}, together with the path structure explained above, allow solving \eqref{DiscreteMSBP} with computational complexity that is linear in $n_s$ and linear in $n_d$. This is particularly nice considering that the primal formulation \eqref{DiscreteMSBP} has exponential complexity.

Notice also that both the formulation and the solution of the MSBP \eqref{DiscreteMSBP} are nonparametric in the sense no statistical parametrization (e.g., mixture of Gaussian, exponential family, knowledge of moment or sufficient statistic) is assumed whatsoever on the joint $\bm{M}_{\rm{opt}}$.

Once the MSB $\bm{M}_{\rm{opt}}$ is computed using \eqref{Moptstructure}-\eqref{MultimarginalSink}, we obtain the distributions $\mu_t$ in \eqref{ComputeConditional} as follows. Let
$$ M_{\rm{opt}}^\sigma := {\rm{proj}}_{\sigma,\sigma+1}(\bm{M}_{\rm{opt}}).$$
For $t\in[t_1,t_{n_s}]$, we compute
\begin{equation}
    \mu_t = \sum_{i=1}^{n_d}\sum_{j=1}^{n_d}[(M_{\rm{opt}}^\sigma)_{i,j}]\delta\Big(\eta-\bigr((1-\lambda)\eta^i(t_\sigma)+\lambda\eta^j(t_{\sigma+1})\bigr)\Big)
\end{equation}
where $t\in[t_\sigma,t_{\sigma+1}]$ and $\lambda:=\tfrac{t-t_\sigma}{t_{\sigma+1}-t_\sigma}\in[0,1]$. Each such $\mu_t$ is a weighted scattered distribution of $n_d^2$ data points, and represents the maximum likelihood distribution $\eta(t)\sim\mu_t$ such that $\eta(t_\sigma)\sim\mu_\sigma$ for all $\sigma\in\llbracket n_s\rrbracket$ and given transport cost $\bm{C}$.

\end{multicols*}


\end{document}